\documentclass[usenatbib,usegraphicx]{mn2e}
\usepackage{amssymb}
\newcommand{\s}{\mathop{\rm s\,}\nolimits}
\newcommand{\m}{\mathop{\rm m\,}\nolimits}

\newcommand{\Ma}{\mathop{\rm Myr\,}\nolimits}

\newcommand{\Hz}{\mathop{\rm Hz\,}\nolimits}
\newcommand{\cm}{\mathop{\rm cm\,}\nolimits}
\newcommand{\km}{\mathop{\rm km\,}\nolimits}

\newcommand{\Mpc}{\mathop{\rm Mpc\,}\nolimits}

\newcommand{\J}{\mathop{\rm J\,}\nolimits}

\newcommand{\eV}{\mathop{\rm eV\,}\nolimits}

\newcommand{\K}{\mathop{\rm K\,}\nolimits}
\newcommand{\W}{\mathop{\rm W\,}\nolimits}

\newcommand{\ryd}{\mathop{\rm ryd\,}\nolimits}

\newcommand{\halfspace}{\hspace{1pt}}
\newcommand{\flux}{\mathop{\mathcal{F\,}\,}\nolimits}
\newcommand{\Lya}{Ly$\alpha$}
\newcommand\HI{{\hbox{H\halfspace$\rm \scriptstyle I$}}}
\newcommand\HII{{\hbox{H\halfspace$\rm \scriptstyle II$}}}
\newcommand\HeI{{\hbox{He\halfspace$\rm \scriptstyle I$}}}
\newcommand\HeII{{\hbox{He\halfspace$\rm \scriptstyle II$}}}
\newcommand\HeIII{{\hbox{He\halfspace$\rm \scriptstyle III$}}}
\newcommand\CIV{{\hbox{C\halfspace$\rm \scriptstyle IV$}}}
\newcommand\SiIV{{\hbox{Si\halfspace$\rm \scriptstyle IV$}}}
\newcommand\Hs{{\rm H}}
\newcommand\HIs{{\rm H\halfspace\scriptscriptstyle I}}
\newcommand\HIIs{{\rm H\halfspace\scriptscriptstyle II}}
\newcommand\Hes{{\rm He}}
\newcommand\HeIs{{\rm He\halfspace\scriptscriptstyle I}}
\newcommand\HeIIs{{\rm He\halfspace\scriptscriptstyle II}}
\newcommand\HeIIIs{{\rm He\halfspace\scriptscriptstyle III}}
\newcommand\LCDM{$\Lambda$CDM}

\newcommand{\fig}{Fig.~\ref}
\newcommand{\figs}{Figures~\ref}
\newcommand{\Fig}{Figure~\ref}
\newcommand{\tab}{Table~\ref}

\newcommand{\sect}{\S\ref}

\newcommand{\eq}{Eq.~\ref}

\newcommand{\expd}[1]{\times 10^{#1}}
\newcommand{\mean}[1]{\langle #1 \rangle}
\newcommand{\vect}[1]{\mathop{\bf #1\,}\nolimits}

\newcommand{\fraction}[2]{\mbox{\scriptsize$^{{#1}\!}/_{\!{#2}}$}}
\newcommand\lsim{~\lower.5ex\hbox{$\buildrel < \over \sim$}~}
\newcommand\gsim{~\lower.5ex\hbox{$\buildrel > \over \sim$}~}
\newcommand{\expectation}[1]{\hbox{$<$}#1\hbox{$>$}}

\ifx\undefined\onequarter
 \newcommand{\onequarter}{\fraction{1}{4}}
\else
 \renewcommand{\onequarter}{\fraction{1}{4}}
\fi
\ifx\undefined\onefourth
 \newcommand{\onefourth}{\onequarter}
\else
 \renewcommand{\onefourth}{\onequarter}
\fi
\ifx\undefined\onethird
 \newcommand{\onethird}{{\fraction{1}{3}}}
\else
 \renewcommand{\onethird}{{\fraction{1}{3}}}
\fi
\ifx\undefined\onehalf
 \newcommand{\onehalf}{{\fraction{1}{2}}}
\else
 \renewcommand{\onehalf}{{\fraction{1}{2}}}
\fi
\ifx\undefined\twothirds
 \newcommand{\twothirds}{\fraction{2}{3}}
\else
 \renewcommand{\twothirds}{\fraction{2}{3}}
\fi
\ifx\undefined\threequarters
 \newcommand{\threequarters}{\fraction{3}{4}}
\else
 \renewcommand{\threequarters}{\fraction{3}{4}}
\fi

\ifx\undefined\onetenth
 \newcommand{\onetenth}{\fraction{1}{10}}
\else
 \renewcommand{\onetenth}{\fraction{1}{10}}
\fi

\title[Reionisation scenarios and the temperature of the IGM]{
       Reionisation scenarios and the temperature of the IGM}
\author[Eric R. Tittley and Avery Meiksin]{
        Eric R. Tittley$^{1}$ and Avery Meiksin$^{1}$
        \thanks{E-mail: ert@roe.ac.uk (ERT); aam@roe.ac.uk (AM)}\\
        $^{1}$SUPA\thanks{Scottish Universities Physics Alliance},
	Institute for Astronomy, University of Edinburgh,
        Blackford Hill, Edinburgh EH9 3HJ, UK}

\shortcites{}

\begin{document}

\date{Accepted . Received ; in original form }
\pagerange{\pageref{firstpage}--\pageref{lastpage}} \pubyear{2007}
\maketitle
\label{firstpage}

\begin{abstract}
We examine the temperature structure of the IGM due to the passage of
individual ionisation fronts using a radiative transfer (RT) code
coupled to a particle-mesh (PM) $N$-body code. Multiple simulations
were performed with different spectra of ionising radiation: a power
law ($\propto \nu^{-0.5}$), miniquasar, starburst, and a time-varying
spectrum that evolves from a starburst spectrum to a power law. The RT
is sufficiently resolved in time and space to correctly model both the
ionisation state and the temperature across the ionisation front. We
find the post-ionisation temperature of the reionised intergalactic
medium (IGM) is sensitive to the spectrum of the source of ionising
radiation, which may be used to place strong constraints on the nature
of the sources of reionisation. Radiative transfer effects also
produce large fluctuations in the \HeII\ to \HI\ number density ratio
$\eta$. The spread in values is smaller than measured, except for the
time-varying spectrum. For this case, the spread evolves as the
spectral nature of the ionising background changes. Large values for
$\eta$ are found in partially ionised \HeII\ as the power-law spectrum
begins to dominate the starburst, suggesting that the large $\eta$
values measured may be indicating the onset of the \HeII\ reionisation
epoch.
\end{abstract}

\begin{keywords}
radiative transfer --
cosmology: diffuse radiation --
cosmology: large-scale structure of Universe --
methods: numerical --
methods: N-body simulations
\end{keywords}

\section{Introduction}
\label{sec.Intro}
The process by which the Universe was reionised is one of the premier
unsolved questions in cosmology. Measurements of the \Lya\ optical
depth of the Intergalactic Medium (IGM) along Quasi-Stellar Object
(QSO) lines-of-sight require the IGM to have been reionised by $z\ga6$
\citep{Becker01}, consistent with recent measurements of the Cosmic
Microwave Background (CMB) by the {\it Wilkinson Microwave Anisotropy
Probe} ({\it WMAP}), although a higher redshift of $z \sim 11$ is
preferred, with a 2-$\sigma$ upper limit of $z\la17$
\citep{Spergel06}. The sources of the reionisation are currently
unknown. The most recent estimates of the numbers of high redshift
QSOs suggest QSOs are too few to have ionised the \HI\ prior to
$z\approx4$, without an upturn in the QSO luminosity function at the
faint end \citep{Meiksin05}. While an adequate supply of ionising
photons is likely produced within young, star-forming galaxies, it has
yet to be conclusively demonstrated that the ionising photons are able
to escape in sufficient numbers to meet the requirements for
reionisation \citep{FLC03,MWK03}. Other, more speculative,
possibilities include pockets of Population III stars (e.g., in young
galaxies or star clusters)\citep{CF06}
or miniquasars \citep{Madau04}.

The post-ionisation IGM temperature may provide a clue to the nature
of the reionisation sources. Low density regions retain a memory of
the post-ionisation temperature \citep{Meiksin94,MR94}. Evidence for
temperatures in excess of the optically thin predictions at high
redshifts ($z>3$) is provided by the widths of the low column density
\Lya\ forest absorbers \citep{MBM01} which are expected to reside in
the underdense regions comprising most of the volume of the Universe
\citep{ZMAN98}.

In the past few years, several groups have implemented numerical
radiative transfer schemes to solve for the reionisation of the IGM
\citep{ANM99,GA01,NUS01,RNAS02,CSW03,Iliev06,WN06}. Most of the
modelling of IGM reionisation has focused on the propagation of
ionisation fronts (I-fronts), with less emphasis given to the accurate
computation of post-ionisation temperatures \citep{Iliev06b}, as the
required calculations are much more computationally demanding
\citep{BMW04,BH06}. Crucial to accurate temperatures is the resolution
of the ionisation fronts, particularly for hard spectrum sources like
young galaxies or Active Galactic Nuclei (AGN). Early estimates of the
post-ionisation IGM temperature were made assuming optically thin
gas. Accounting for radiative transfer within the ionisation fronts
produces much higher temperatures.

In this paper, we focus on the thermal properties of the post-ionised
IGM. We investigate the role played by the spectral shape of the
dominant ionising sources on the temperature of the post-ionised IGM,
considering both soft sources like galaxies and hard like QSOs. Rather
than performing a full reionisation simulation with multiple sources,
which is still beyond the current resources of numerical computations,
we limit the computations to a single ionisation front sweeping across
the box, as would occur prior to the time of the complete overlap of
ionisation fronts. Although this approximation will not allow us to
compute precise values for the ionisation fractions of hydrogen and
helium for comparison with measurements, as these levels will be reset
by the subsequent total photoionisation rates after I-fronts overlap,
it does permit a quantitative evaluation of the {\it temperature
structure} after complete reionisation. This is because, once ionised,
the temperature is insensitive to the total intensity or shape of the
ionising background \citep{Meiksin94, HG97}, which will
readjust both due to the overlapping of I-fronts and the evolution of
the sources and photoelectric optical depth of the IGM.

The simulations are performed with a radiative transfer code coupled
to an $N$-body code to model the evolution of the IGM. In
\sect{sec.Method} the algorithm and the numerical code are
described. The details of the simulation volume and sources are given
in \sect{sec.Simulations}. Results of the simulations are provided in
\sect{sec.Results} and discussed in \sect{sec.Discussion}.

\section{Method}
\label{sec.Method}
The simulation code, {\tt PMRT\_mpi}, is the merger of a Lagrangian
particle-mesh (PM) code \citep{MWP99}, and a grid-based radiative
transfer (RT) code which are modularly independent. We perform the RT
on the evolving gas density field, as computed by the PM code. Full
hydrodynamical simulations have shown that the gas density closely
traces the dark matter density down to the Jeans scale ($\sim100$~kpc)
\citep{Cen94,ZMAN98}. Statistical comparisons between the resulting
\Lya\ forest properties show good agreement at the 10 per cent level
\citep{MW01,Viel02}.

The gas density is taken as a constant fraction ($\Omega_b / \Omega_m$) of the
total density. This is a fair approximation since baryons trace dark matter
\citep{ZMAN98} and in simulations with gas, the gas to dark matter ratio doesn't vary
more than a factor of 30 percent over the over density range $\delta=0.1$ to
1000 \citep{TC00}. The I-fronts propagate much faster than the sound speed, so
that the pressure-response of the gas has only a small effect on the
reionisation. The PM code evolves the density field assuming all mass is
collisionless, interacting only gravitationally.

The density field is determined from the particle
distribution by griding the particles on to a mesh. To avoid low-count
artefacts in low-density regions while not sacrificing information in
dense regions, the density field is separated into two fields,
$\rho(\vect{r}) = \rho_{lo}(\vect{r}) + \rho_{hi}(\vect{r})$, and the
low-density field, $\rho_{lo}(\vect{r})$, is convolved with a Gaussian
of radius two grid cells. A threshold cell count ($N_{thresh}=5$ for
these simulations) discriminates $\rho_{lo}(\vect{r})$ from
$\rho_{hi}(\vect{r})$. In dense regions,
$\rho_{lo}(\vect{r})=N_{thresh}$ and $\rho_{hi}(\vect{r})=
\rho(\vect{r})-N_{thresh}$, while in low-density regions,
$\rho_{lo}(\vect{r})=\rho(\vect{r})$. The smoothed field is then
$\rho'(\vect{r}) = \rho_{lo}(\vect{r})\otimes g(\vect{r}) +
\rho_{hi}(\vect{r})$ where $g$ is the Gaussian smoothing kernel.

\subsection{Radiative transfer}
\label{sec.RT}
The RT code uses a probabilistic method which is based on the
photon-conserving algorithm of \citet{ANM99}, extended to include
helium by \citet{BMW04} who applied the RT algorithm to a density
field frozen in the comoving frame. For convenience, we provide
details of the RT code here.

The rates of change of the ionisation-state populations due to
ionisations and recombinations are given by:
\begin{eqnarray}
\label{eq.IonisationStates}
\dot{n}_\HIs   &=& n_e n_\HIIs \alpha_\HIIs - n_\HIs \Gamma_\HIs \\
\dot{n}_\HIIs  &=& -\dot{n}_\HIs  \nonumber \\
\dot{n}_\HeIs  &=& n_e n_\HeIIs \alpha_\HeIIs - n_\HeIs \Gamma_\HeIs \nonumber \\
\dot{n}_\HeIIs &=& - \dot{n}_\HeIs - \dot{n}_\HeIIIs \nonumber \\
\dot{n}_\HeIIIs&=& n_\HeIIs \Gamma_\HeIIs - n_e n_\HeIIIs \alpha_\HeIIIs \nonumber
\end{eqnarray}
where $\alpha_i$ is the recombination coefficient from species $i$ and $\Gamma_i$ is the photoionsation rate. Number-density changes, $\dot{n}_i$, due to cosmological evolution are accounted for by carrying only the ionisation fractions ($f_\HIs = n_\HIs/n_\Hs$ and similar for helium fraction $f_\HeIs$, $f_\HeIIs$, and $f_\HeIIIs$) between PM iterations.

\begin{table*}
\caption{Recombination ($\alpha_i$ [$\m^3 \s^{-1}$]) and recombination cooling ($\beta_i$ [$\J \m^3 \s^{-1}$]) coefficients.}
\begin{tabular}{rl}
\hline
 $\alpha_\HIIs$  & $2.065\expd{-17} T^{-\onehalf} \left(6.414 - \frac{1}{2}\ln\, T
                   + 8.68\expd{-3} T^\onethird \right)$ \\
 $\alpha_\HeIIs$ & $3.294\expd{-17} \left\{\left(\frac{T}{15.54}\right)^\onehalf
                  \left[1+\left(\frac{T}{15.54}\right)^\onehalf \right]^{0.309}
		  \left[ 1+\left(\frac{T}{3.676\expd{7}}\right)^\onehalf
		        \right]^{1.691}
		  \right\}^{-1}
                  +1.9\expd{-9} 
	          \left( 1 + 0.3 e^{\frac{-9.4\expd{4}}{T}} \right)
	          e^{\frac{-4.7\expd{5}}{T}}  T^{-\fraction{3}{2}}$ \\
 $\alpha_\HeIIIs$ & $8.260\expd{-17} T^{-\onehalf} \left( 7.107 
                     - \frac{1}{2}\ln\,T
                    + 5.47\expd{-3} T^\onethird \right)$ \\
\hline
 $\beta_\HIIs$   & $2.851\expd{-40} T^{\onehalf} \left(5.914 -\frac{1}{2}\ln\, T
                   + 0.01184 T^{\onethird} \right)$ \\
 $\beta_\HeIIs $ & $1.55\expd{-39} T^{0.3647}
                 +1.24\expd{-26} 
	          \left( 1 + 0.3 e^{\frac{-9.4\expd{4}}{T}} \right)
	          e^{\frac{-4.7\expd{5}}{T}}  T^{-\fraction{3}{2}}$ \\
 $\beta_\HeIIIs$ & $1.140\expd{-39} T^\onehalf \left( 6.607 - \frac{1}{2}\ln\, T
                   + 7.459\expd{-3} T^\onethird \right)$ \\
\hline
\label{table.RecombCoef}
\end{tabular}
\end{table*}

The forms for the recombination coefficients $\alpha_\HIIs(T)$ and
$\alpha_\HeIIIs(T)$ are derived from the generalised hydrogenic case A form
given in \citet{Seaton59}.  For $\alpha_\HeIIs(T)$, the radiative term
is from \citet{VF96} while the dielectronic term is adopted from
\citet{AP73}.  The recombination coefficients are provided in
\tab{table.RecombCoef}.

\begin{table}
\caption{Ionisation cross-section parameters used in \eq{eq.CrossSection}.}
\begin{tabular}{rllll}
 & $\sigma_T [\m^2]$ & $\nu_T$ [Hz] & $\beta$ & $\s$ \\
 \hline
 $\sigma_\HIs$   & $6.30\expd{-22}$ & $3.282\expd{15}$ & 1.34 & 2.99 \\
 $\sigma_\HeIs$  & $7.83\expd{-22}$ & $5.933\expd{15}$ & 1.66 & 2.05 \\
 $\sigma_\HeIIs$ & $1.58\expd{-22}$ & $1.313\expd{16}$ & 1.34 & 2.99 \\
 \hline
\label{table.CrossSection}
\end{tabular}
\end{table}

The photoionisation rate per particle, $\Gamma$, for each species is dependent on the local mean intensity of radiation, $J_\nu$, and the ionisation cross section, $\sigma$, for the species by
\begin{equation}
\Gamma = 4\pi \int_{\nu_T}^{\infty} \frac{J_\nu}{h\nu} \sigma_\nu d\nu
\label{eq.Gamma}
\end{equation}
where $\nu_T$ is the threshold frequency for ionisation, which differs for each
species.
In the plane wave approximation used here, the local mean intensity is simply the radiation field incident on the volume attenuated by the cumulative optical depth,
\begin{equation}
4\pi J_\nu = \frac{L_\nu}{4\pi R_0^2} e^{-\tau_\nu}
\end{equation}
where $L_\nu$ is the luminosity of the source per frequency, $R_0$ is the distance of the source from the volume, and $\tau_\nu$ is the cumulative optical depth which, at distance R from the source, is:
\begin{equation}
\tau_\nu(R) = \sigma_\HIs N_\HIs + \sigma_\HeIs N_\HeIs + \sigma_\HeIIs N_\HeIIs
\end{equation}
where the cumulative column depth from the edge of the box to $R$ is
\begin{equation}
N_i(R) = \int_{R_0}^{R}n_i(r) dr .
\end{equation}
The ionisation cross sections are approximated using the form given in \citet{Osterbrock89},
\begin{equation}
\label{eq.CrossSection}
\sigma(\nu) = \sigma_T \left[ \beta
\left(\frac{\nu}{\nu_T}\right)^{-s} +
(1-\beta)\left(\frac{\nu}{\nu_T}\right)^{-s - 1}\right] .
\end{equation}
The parameters for the various species are listed in \tab{table.CrossSection}.
The reionisation is further restricted by imposing a check on the position
of the light front to ensure gas is not ionised too early. Neglecting
this effect can lead to an over-estimate of the temperature \citep{MMR97}.

The equations governing the thermal state of the gas are used in their entropic form.  The entropy is parametrized by the function
\begin{equation}
\label{eq.Entropy}
S \equiv \frac{P}{\rho^\gamma},
\end{equation}
where $P$ is the pressure, $\rho$ is the gas density, and $\gamma = 5/3$ is the adiabatic index for a monatomic gas. It follows from \eq{eq.Entropy} that
\begin{equation}
\dot{S} = (\gamma-1)\rho^{-\gamma}(G-L),
\end{equation}
where $G$ and $L$ are the thermal gain and loss functions per volume, and
\begin{equation}
T = \frac{\mu m_u}{k} S \rho^{\gamma-1} 
\label{eq:TS}
\end{equation}
where $\mu$ is the mean molecular weight of the gas, $m_u$ is an atomic
mass unit, and $k$ is the Boltzmann constant.

Heating is provided by the excess energy above the ionisation threshold $h\nu_T$ of the ionising photons. For a single species of density $n$,
\begin{equation}
\label{eq.IonisationHeating}
G = 4\pi n \int_{\nu_T}^{\infty}\frac{J_\nu \sigma_\nu}{\nu}(\nu-\nu_T)d\nu .
\end{equation}
Since all species contribute, the total heating rate is
\begin{equation}
G = G_\HIs + G_\HeIs + G_\HeIIs.
\end{equation}

Cooling is provided by recombinations, collisional excitation of the excited levels in neutral hydrogen, and inverse Compton scattering off cosmic microwave background (CMB) photons. As for heating, all species contribute to cooling, giving the total cooling rate
\begin{equation}
L = L_\HIIs + L_\HeIIs + L_\HeIIIs + L_{eH} + L_C.
\end{equation}

For a single species, recombinations radiate the electron energy $\sim kT$ as photons at the rate $L_i = n_e n_i \beta_i(T)$. From \HII\ and \HeIII , the recombination cooling coefficients $\beta_i(T)$ are, as $\alpha_i(T)$, from \citet{Seaton59}. For the \HeII\ radiative term we use the expression in \citet{Black81} while we combine the approximation $\beta_\HeIIs = 3 \ryd \alpha_\HeIIs$ \citep{GT70} with the second $\alpha_\HeIIs$ term listed in \tab{table.RecombCoef} \citep{AP73} for the dielectronic component. The total recombination cooling coefficients are provided in \tab{table.RecombCoef}.

For the cooling rate from collisional excitation of \HI\ we adopt the approximation of \citet{Spitzer78}:
\begin{equation}
L_{eH} = 7.3\expd{-32} \J \m^3 \s^{-1} n_e  n_\HIs  e^{-118400 / T}.
\end{equation}
We note that for the temperatures relevant here, cooling losses due to
collisional ionisation of hydrogen and collisional excitation and
ionisation losses from helium are negligible.

Finally, Compton scattering off CMB photons cools the gas at the rate \citep{Peebles68}
\begin{equation}
L_C = \frac{4 \sigma_T a k}{m_e c} n_e T^4_{CMB}(z) \left[T - T_{CMB}(z)\right],
\end{equation}
where $\sigma_T$ is the Thomson cross section, $a$ is the radiation density constant, $m_e$ is the electron mass, $c$ is the speed of light, and $T_{CMB}$ is the temperature of the microwave background.

In the current implementation, the simulations do not solve for the overlapping of ionisation fronts. Rather, they describe the passage of the first ionisation front across a neutral region.
Reionisation will also be affected by the hydrodynamical response of the gas, which is not included here. For low to moderately overdense systems, this has only a moderate effect on the statistical properties of the resulting absorption systems \citep{MW01}. The hydrodynamical effects are more important for reionisation in denser structures like minihaloes, which can be optically thick at the Lyman edge. Ionisation heating results in an overpressure that drives the gas out of the minihaloes, reducing their densities and making them less effective at slowing the ionisation fronts. Estimates based on the photon consumption rate suggest the role of these systems in slowing the fronts is small \citep{Ciardi06}.

Another simplification is the absence of diffuse radiation.  Radiative recombinations produce a diffuse radiation field throughout the ionised region. Besides the recombination rate, the intensity depends on the amount of clumping of the gas. Estimates range from a boost in the ionisation rate by an additional 10--40 per cent. \citep{MM93, HM96}. After the gas has been ionised, the diffuse field may be accounted for by rescaling the overall radiation level, as the intensity of the radiation field has a negligible effect on the post-ionisation gas temperature. Radiative recombinations will also contribute to the reionisation process itself. The contribution, however, is generally negligible for both hydrogen and helium reionisation \citep{SG87, MO90, MM93, MM94}. An
exception is in dense regions in a scenario in which the \HeIII\ I-front precedes the \HII\ I-front. In this case, assuming all photons produced by helium recombinations to the ground state of \HeII\ are locally absorbed by neutral hydrogen atoms, the time to ionise the \HI\ is
\begin{eqnarray}
t_{\rm He-H}&\approx&\frac{n_{\rm H}}{5n^2_{\rm He}\alpha_1^{\rm He}}\\
&\approx&7.7\times10^{20}\,{\rm s}\left(\frac{\Omega_b h^2}{0.020}\right)^{-1}
\frac{T_4^{0.5}}{1+\delta}(1+z)^{-3}\nonumber,
\label{eq:tHeH}
\end{eqnarray}
where $\alpha_1^{\rm He}$ is the radiative recombination rate to the
ground state of \HeII, $T_4$ is the gas temperature in units of $10^4$~K, and $\delta$ is the local fractional overdensity. An allowance has also been made for 1.5 additional secondary electrons produced per photoionisation \citep{Shull79}. Compared with a Hubble time of $t_{\rm Hub}\approx14\,{\rm Gyr}(1+z)^{-3/2}$, the ionisation time goes as
\begin{equation}
\frac{t_{\rm He-H}}{t_{\rm Hub}}\approx 640T_4^{0.5}\frac{1}{(1+z)^{3/2}
(1+\delta)}.
\label{eq:tHeH-Hub}
\end{equation}
For the range of source turn-on redshifts considered ($8<z<20$), except in very overdense regions, this ratio will be large and hydrogen will remain largely neutral in the interval between helium and hydrogen ionisation. The effect on the temperature will therefore be small, except for very overdense regions. After reionisation, the post-ionisation temperature will be accurately computed even in the highly overdense regions since the temperature is set by the balance between ionisation heating and radiative cooling with no memory of the reionisation history retained \citep{Meiksin94}.

Heating by x-rays is automatically included, but not the effect of
secondary electrons. The secondary electrons have a negligible effect
on the post-ionisation temperature of the gas \citep{MMR97}. Ahead of
the I-front, however, the heating rates will be overestimated by a
factor of 2--5, depending on optical depth to the ionising photons
(which determines their mean energies)
\citep{SS85}. While the tendency to approach thermal
equilibrium reduces the impact of the secondaries on the final
temperature, the gas temperature ahead of the I-front is somewhat
reduced when the effect of the secondaries is included, as discussed
in the next section. Because of the excessive computational expense
involved in including the effect of secondaries, we neglect their role
on the pre-ionised gas temperatures, but note the effect of
secondaries must be included for applications for which the
pre-ionised gas temperatures are relevant, as for instance the 21cm
signature of the gas at high redshifts \citep{MMR97}.

None of the above simplifications affects our findings presented in
this paper, which concentrates on the post-ionisation thermodynamic
properties of the IGM as it relates to the \Lya\ forest.

\subsection{Implementation}
\label{sec.Implementation}
To solve the RT equations, we use the probabilistic
formulation described in Abel et al. (1999), as extended to
multiple species by Bolton et al. (2004). In the probabilistic formulation,
the ionisation rate per unit volume due to an incident photon flux per frequency interval, $\flux_\nu$, is given by
\begin{equation}
n \Gamma = \frac{1}{dr}\int_{\nu_0}^\infty \flux_\nu \sigma_\nu n dr d\nu .
\label{eq.IonisationsPerLength}
\end{equation}
is approximated as
\begin{equation}
n \Gamma = \frac{1}{dr}\int_{\nu_0}^\infty \flux_\nu
           \left( 1-e^{-\tau_\nu^{\rm cell}} \right) d\nu
\label{eq.IonisationsPerLength_tau}
\end{equation}
where $\tau_\nu^{\rm cell}=\sigma_\nu n dr$ is the optical
depth of a single cell of width $dr$. This formulation conserves
photons and does not lead to excessive ionisations through cells
for which $\tau_\nu^{\rm cell}>1$.
For multiple species, the probability of absorption is spread among the various species and the rate of ionisation must be known for each species. \citet{BMW04} describes the split for absorption by \HI , \HeI , and \HeII . Using their shorthand $p^i = e^{-\tau_\nu^i}$ and $q^i = 1 - e^{-\tau_\nu^i}$ where the $\tau_\nu^i$'s are within the cell (as opposed to cumulative), the probabilities of absorption by each species are
\begin{eqnarray}
P_{abs}^\HIs   &=& q^\HIs p^\HeIs p^\HeIIs
                   \left(1 - e^{-\tau_\nu^{\rm cell}} \right)/D \\
P_{abs}^\HeIs  &=& q^\HeIs p^\HIs p^\HeIIs
                   \left(1 - e^{-\tau_\nu^{\rm cell}} \right)/D \\
P_{abs}^\HeIIs &=& q^\HeIIs p^\HIs p^\HeIs
                   \left(1 - e^{-\tau_\nu^{\rm cell}} \right)/D
\end{eqnarray}
with $D=q^\HIs p^\HeIs p^\HeIIs + q^\HeIs p^\HIs p^\HeIIs + q^\HeIIs p^\HIs p^\HeIs$ and $\tau_\nu^{\rm cell} = \tau_\nu^\HIs + \tau_\nu^\HeIs + \tau_\nu^\HeIIs$. Equation~\ref{eq.IonisationsPerLength} with $\sigma_\nu n dr$ replaced by the $P_{abs}^i$'s above is the form implemented in {\tt PMRT\_mpi}. We recast \eq{eq.IonisationHeating} for each species in a similar manner, to prevent excessive heating.

Both the PM and RT modules were coded in C and parallelised using a message passing interface (MPI) (the PM component previously so \citep[see][]{MWP99}). MPI is suited to work on distributed memory systems, but works equally well on shared memory systems, albeit not with optimal memory use. The RT module consumes most of the processing time. In the RT module, each LOS is processed by a single computer processor element (PE). In general the time spent by each process on a LOS scales linearly with the resolution along the LOS. In regions of high optical depth, refinements are generated which split the cell into low-optical-depth sections. The number of refinements is a factor in the PE load for a LOS. Load imbalance is possible because the time to process a LOS varies for each LOS, depending on the amount of structure. Load imbalance is alleviated by dynamic (first-come-first-serve) allocation of LOSs to processes. The master node performs the allocation and, consequently, has a significantly lower mean load. The load imbalance with the master node can be reduced by either 1) having one physical processor unit execute both the master and a slave process or 2) using many processor units, so that instead of one-of-few being underused, one-of-many is underused and less time is spent in a non-load-balanced state.

In addition to accurate modelling of the ionisation front, special attention is given to computing accurate post-photoionisation temperatures. This requires sufficient resolution in time and space to correctly model both the ionisation structure of the ionisation front and the temperature across it. This is particularly important when modelling reionisation by sources with hard photons like QSOs.  The time step for the radiative transfer is computed, independent of the PM time step, from the cooling, ionisation, recombination, and density-change time scales.  The minimum of these time scales is multiplied by a factor of 0.2, found through convergence tests using a uniform medium containing \HeIII\ and \HII\ fronts. Convergence is considered established if the maximum error in the temperature along the LOS (compared with a fiducial run using a factor of 0.1) is less than half a per cent.  High spatial resolution at the front is guaranteed by the use of refinements.  Any cell in a line of sight (LOS) which is optically thick at the Lyman edge ($\tau > 1$) is split into a sufficient number of slices such that $\tau \le 1$ in each. Convergence tests were similarly used to establish this criterion by comparing with a fiducial run with $\tau \le 0.05$ per slice. Refinements are normally only generated at the ionisation front.

Integration over frequency of the ionisation (\eq{eq.Gamma}) and heating (\eq{eq.IonisationHeating}) functions is performed using Gauss-Legendre quadrature over the intervals $\nu_\HIs$ to $\nu_\HeIs$ and $\nu_\HeIs$ to $\nu_\HeIIs$ and Gauss-Laguerre quadrature for the interval $\nu_\HeIIs$ to $\infty$ where $\nu_i$ is the ionisation threshold for species $i$. The integration parameters for each of these intervals was tailored to each of the input spectra using convergence tests similar to those described in the previous paragraph.

Of course, the effort required to get the temperatures correct comes at the expense of computer cycles. More than 20000 PE hours were required for the simulations presented here. {\tt PMRT\_mpi} was run in parallel on 8 or 16 PEs drawn from a dedicated cluster of IBM OpenPower 720s or a local cluster of Linux boxes of mixed type.

Validation of the code was accomplished by the use of problems with
known solutions and comparison with results produced by an independent
code. The analytic solution to the position of a hydrogen ionisation
front in a gas of uniform density irradiated by a monochromatic source
with photon energy $h\nu = 13.6 \eV$ (producing no heating) is easily
derivable \citep[][for example]{Iliev06b}. We simulated the passage of
an I-front through a gas with $n_H = 10^{-3} \cm^{-3}$ and $T =
10^4\K$ ionised by a source producing photons at a rate of $5\expd{48}
\s^{-1}$ (Test 1 of \citet{Iliev06b}). Over a period of 500 Myr (a few
times the recombination time scale), the error in the position of the
front was never more than 6 per cent. There is no analytical solution
for the gas temperature if irradiated by a non-monochromatic
source. However, other simulations exist to which we could compare.
One of us (AM) has a radiative transfer code which implements an
altogether different method with more physics than we included in {\tt
PMRT\_mpi} \citep{MMR97}. For an $\alpha=0.5$ power-law spectrum
ionising a uniform IGM at the mean baryon density, this code
(with the effect of secondary electrons switched off) produces
post-ionisation gas temperatures of about 18,000~K to 20,000~K at
$z=7$. Including the effect of secondary electrons was found to
reduce the post-ionisation temperature by about 200~K while the
temperature in the gas ahead of the I-front, where helium ionisation
releases large numbers of electrons, was reduced by about 1000~K. Our
{\tt PMRT\_mpi} code produces post-ionisation gas temperatures of
16,000~K to 23,000~K at this epoch. {\tt PMRT\_mpi} generates the
correct ionisation structure and temperatures that agree with both
expectations from observations and the result of an independent
code. We are confident that given the physics included, the {\tt
PMRT\_mpi} code is producing correct results.

\section{Simulations}
\label{sec.Simulations}

We use {\tt PMRT\_mpi} to simulate the passage of an ionisation front
produced by sources with a variety of characteristic spectra. Each
I-front computed may be considered to result from a single source with
the adopted spectrum, or a group of several neighbouring sources with
individual I-fronts that have already overlapped, producing a single
advancing I-front through the IGM. Because we treat each line-of-sight
separately, the resulting temperature distributions may also be
considered as produced by distinct sources, and the resulting
temperature statistics an ensemble average over the IGM.

In the following, we describe the sources of ionising radiation
(\sect{sec.SimSources}), estimate the expected characteristic scales
of their ionisation fronts at the reionisation epoch
(\sect{sec.Ifront}), describe the volume of space simulated
(\sect{sec.SimVolume}) and demonstrate that the results have converged
with resolution and are not dominated by cosmic variance
(\sect{sec.SimVariance}).

\subsection{Sources}
\label{sec.SimSources}
\begin{figure}
\includegraphics{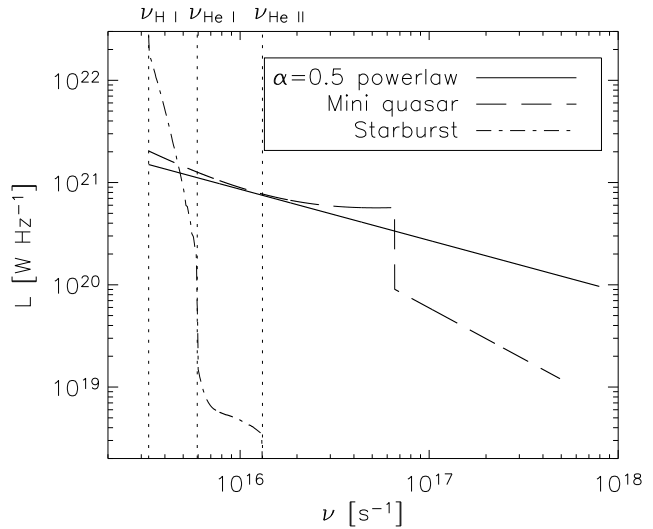}
\caption{Source spectra. The ionisation thresholds for \HI, \HeI, and
\HeII\ are indicated.}
\label{fig.spectra}
\end{figure}
The source spectra (\fig{fig.spectra}) were selected to emulate
candidate reionisation sources. For the power law, the luminosity is
given by $L(\nu) \propto \nu^{-\alpha}$ where $\alpha = 0.5$, which
corresponds to a hard quasar/QSO/AGN spectrum. The miniquasar
spectrum of \citet{Madau04} is given by $L(x) \propto x^{\onefourth} +
8x^{-1}$, $1 \le x \le 20$; $P(x) \propto 8x^{-1}$, $x \ge 20$ where
$x \equiv \nu/\nu_0$. The starburst spectrum was produced by {\tt
P\'EGASE}\footnote{\tt http://www2.iap.fr/users/fioc/PEGASE.html}
\citep{FR97} for a galaxy 30 Myr after a burst of Pop III (zero
metallicity) star formation. The starburst spectrum has an effective
spectral index of $\alpha_{\rm eff}=7.4$ just above the Lyman edge.
The hybrid model begins with a starburst spectrum, and between $z=5$
and $z=4$ evolves into an $\alpha = 0.5$ power law, mimicking a radiation
field in which a QSO source dominates.

The order of propagation of the \HII\ and \HeIII\ ionisation fronts is
different for the different sources. For a time-invariant spectrum, the
\HeIII\ I-front precedes the \HII\ I-front if the spectrum is hard
enough, specifically if $\alpha_{\rm eff} < 1.8$. The condition is met
for both the power-law and miniquasar spectra and, indeed, the \HeII\ 
was ionised prior to \HI\ in those simulations. The starburst spectrum
certainly fails the condition as it has negligible intensity above
$\nu_\HeIIs$. The hybrid spectrum, in which a spectrum dominated by
star bursts evolves into one dominated by a power law, leads to the
\HII\ I-front preceding that of \HeIII .

To ensure comparable results, all spectra were normalised to produce an incident hydrogen ionising
flux of $1.5\expd{7} (1+z)^2 \s^{-1} \m^{-2}$. This corresponds to a
typical flux level driving an ionisation front, as detailed in \sect{sec.Ifront}. For example, it
corresponds to a $L[\nu_\HIs] = 1.5\times 10^{21} \W \Hz^{-1}$ QSO
source with $\alpha = 0.5$ at a comoving distance of $5 \Mpc$. The
flux is adequate to ionise the full simulation volume by $z=3$. We fix
the incident flux level to a common level to ensure that differences
in the post-reionisation temperature are due to the differences in the
incident spectra rather than the time scale for reionisation.  We also
explore the effect of source turn-on redshift by performing, for each
source, simulations with the source turning on at redshifts of $z_{\rm
on}=8$, 12, and 20. These were selected to cover the range of
redshifts limited by the {\it WMAP} and QSO observations. Because of
the common incident flux normalisation, these do not correspond to
different reionisation scenarios in which the reionisation of the
Universe completes at different epochs, as the reionisation histories
are nearly identical for $z<8$. It is simply a device that
allows us to ensure the results at late times are insensitive to the
assumed turn-on redshift of the sources.

The ionising radiation is projected as parallel rays normal to the
surface of the simulation volume. This configuration has the
computational advantage of allowing all radiative calculations for a
given ray to be done in a single column, independent of neighbouring
columns.  Since no information about the thermal state of the gas is
carried with the particles as they move from one cell to another, each
column can be interpreted as an individual line of sight which is
independent of any other line of sight. We select 256 lines of sight
arranged in a plane to assist in visually interpreting the passage of
the front while providing enough lines of sight to deal with cosmic
variance. The radiation field takes the plane-wave approximation and
is not geometrically diluted by $r^{-2}$. A limited number of
simulations performed with the radiation field geometrically diluted
produced results similar to those without the dilution factor.

\begin{table}
\caption{Identifiers for the various models used in this paper.}
\begin{tabular}{rccc}
\hline & \multicolumn{3}{c}{$z_{\rm on}$}\\ Model & 8 & 12 & 20 \\
            \hline Power Law & PL08 & PL12 & PL20 \\ Miniquasar &
            MQ08 & MQ12 & MQ20 \\ Starburst & SB08 & SB12 & SB20 \\
            Hybrid & HY08 & HY12 & HY20 \\ \hline
\label{table.identifiers}
\end{tabular}
\end{table}

For brevity, the labels listed in \tab{table.identifiers} will be used
henceforth to identify the various simulations with their combination
of model spectrum and source turn-on redshift.

\subsection{Characteristic sizes of ionised regions}
\label{sec.Ifront}

A key effect for which radiative transfer is required is the
modification of the source spectrum incident on a region by
intervening gas, including shadowing by dense knots. The amount
of modification depends on the amount and nature of the intervening
gas. In this section, we estimate the typical distances expected
between a source and the I-front it produces, and show that our
simulations match the distances for a wide range of plausible sources.
This is important, because the no-overlap approximation we make is
only valid provided the I-fronts do not extend further than the
characteristic size of an I-front at the time of complete
reionisation, when the porosity\footnote{The porosity of the ionised
gas is the product of the source number density and ionisation volume
produced by an individual source.} of the ionised gas in the IGM
crosses unity.

At the time of complete reionisation, the mean space density of the
sources is the inverse of the characteristic volume of the ionised gas
bubbles just before their complete percolation. For any given
characteristic luminosity per source, the source space density may be
estimated from the total emissivity of all the sources just prior to
complete reionisation. We adopt an estimate for the emissivity based
on the constraints imposed by the measured IGM \Lya\ optical depth up
to $z\approx6$ in a $\Lambda$CDM cosmology
\citep{MW03,MW04,Meiksin05}. At $z=6$, the rate of \HI\ ionising
photons per comoving volume was found to be $\dot
n_S\approx4\times10^{50}\alpha_S^{-1}h{\rm
s^{-1}\,Mpc^{-3}}$\citep{Meiksin05}, where $\alpha_S$ is the spectral
index of the ionising sources at the \HI\ Lyman edge and $h$ is
defined by the Hubble parameter ($H_0 = 100 h \km \s^{-1} \Mpc^{-1}$,
where $H_0$ is the present day value of the Hubble parameter). For a
source producing ionising photons at the rate $\dot N_\gamma$, the
corresponding volume of ionised gas at the time of overlap is then
$V_I\approx \dot N_\gamma/\dot n_S$. The characteristic comoving size of an
I-front at $z=6$ is then (for $h=0.71$),
\begin{equation}
R_I\approx\left(\frac{3}{4\pi}\frac{\dot N_\gamma}{\dot n_S}\right)^{1/3}\approx9
\alpha_S^{1/3}\left(\frac{\dot N_\gamma}{10^{54}\,{\rm s^{-1}}}\right)^{1/3}\,{\rm Mpc}.
\label{eq:RI}
\end{equation}
This scale agrees with estimates made by others. \citet{WL04} determined
at overlap the ionised regions are constrained to have scales of 5 to 50
$h^{-1}$ Mpc comoving.
Similar scales have been inferred using semi-analytical reionisation approaches
\citep{FO05,FMH06,CC07}.

In the following sections, we estimate the characteristic I-front size at the time of percolation
for Lyman-break galaxies, low luminosity AGN, and a few other plausible sources.

\subsubsection{Lyman-break galaxies}

Lyman-break galaxies discovered at $z>6$ have been suggested as
principal sources of the reionising photons, depending on the
uncertain escape fraction of ionising radiation from the galaxies, the
nature of the dominant population of stars in the galaxies, and the
clumping factor of the IGM \citep{Bunker04,
SFP04,YW04}. The sources have
characteristic UV (1500~A) luminosities of
$L_\nu\approx10^{28}-10^{29}\,{\rm erg\,s^{-1}\, Hz^{-1}}$
\citep{Steidel99, Bunker06}. The corresponding
rate of ionising photon production depends on the conversion rate
between ionising photons and UV light. Using the values for Population
III, Population II and Population I (solar abundance) stars
\citep[e.g.][]{SFP04}, we obtain $\dot
N_\gamma=(4.8,\,1.0,\,0.17)\times10^{55}\rm s^{-1}$, respectively.  Allowing
for an escape fraction $f_{\rm esc}$, \eq{eq:RI} gives for the
characteristic comoving I-front radius at the time of overlap
$R_I\approx15(f_{\rm esc}/0.1)^{1/3}\,{\rm Mpc}$ for
$\alpha_S\approx2-7.4$ \citep{Meiksin05}. The uncertainty in the
ionising photon conversion rate and escape fraction corresponds to an
additional factor of uncertainty of about two in the size.  Addressing the size
of the ionised bubbles in which QSOs reside prior to turn-on, \citet{AA07} show
that galaxies ionised regions on the order of tens of comoving Mpc
shortly before overlap.

It is possible the IGM was reionised by even rarer, more massive and
more luminous systems, such as massive post-starburst galaxies
\citep{Panagia05}. Allowing for Population III stars and a
high escape fraction, typical I-fronts at the time of overlap could
approach comoving radii of $\sim30-50$~Mpc.

\subsubsection{Active Galactic Nuclei}

The number counts of luminous QSOs fall short
of the amount required to reionise the IGM \citep{YW04,
Meiksin05}. Low luminosity AGN, however, are plausible sources of
non-stellar ionising radiation \citep{RO04}. A population of galaxies harbouring
$10^{6.5}{\rm M_\odot}$ black holes shining at their Eddington
luminosities would require a space density of
$n_g\sim2\times10^{-4}{\rm Mpc^{-3}}$ to reionise the IGM by $z=6$
\citep{Meiksin05}. The corresponding characteristic comoving I-front
radius at the reionisation epoch is then $R_I\approx
n_g^{-1/3}\approx17$~Mpc, comparable to that expected for Lyman-break
galaxies.

It is likely that the sources that ionised \HeII\ to \HeIII\ were
QSOs, as few other objects produce an adequate supply of sufficiently
hard photons.  If the QSOs were low luminosity QSOs, as above, then
the above estimate for the \HII\ fronts would apply to the \HeIII\ 
ionising fronts as well. If the rarer luminous QSOs dominated the
reionisation of \HeII, the characteristic I-fronts at the time of
overlap would be even larger. For a comoving space density at $3<z<5$
of $10^{-7}-10^{-6}\,{\rm Mpc^{-3}}$ \citep{Meiksin05}, the
characteristic comoving sizes would be $100-200\,{\rm Mpc}$.

\subsubsection{Other sources}

It is possible that more common structures ionised the IGM, such as
collapsed systems smaller than Lyman-break galaxies which merged into
larger systems. In the simulations of \citet{Gnedin00},
reionisation is dominated by systems with total stellar masses
exceeding $3\times10^8\,M_\odot$. As only a few such systems are
responsible for ionising the gas in a $4h^{-1}$~Mpc comoving volume,
the characteristic I-front size at overlap is a few comoving Mpc. Our
models are just marginally applicable to such sources, but more
relevant to the rarer larger I-fronts that would occur prior to
overlap than the smaller ionised regions.

In the miniquasar model of \citet{KM05}, an individual
source does not ionise hydrogen much beyond a comoving distance of
20~kpc. In this case, the characteristic I-fronts are much smaller than
those we model. Our miniquasar source spectrum corresponds to luminosities
much larger than those considered by \citet{KM05}. Nevertheless, we adopt the miniquasar model as a physically motivated alternative hard spectral shape. As we show later,
the results are nearly indistinguishable from the pure power-law case.

A more exotic source of ionising radiation is some form of decaying particle.
Various hypotheses have been suggested. Our results are not relevant to a
scenario in which such particles were responsible for the reionisation of
the IGM, as the sources are completely localised and ubiquitous.

\subsection{Simulation volume}
\label{sec.SimVolume}
All simulations were performed in a $(25 h^{-1}\Mpc)^3$ comoving volume. This
is large enough to include the typical pre-overlap I-front for the models
discussed in \sect{sec.Ifront}.
A \LCDM\ model was assumed \citep{Spergel03}, with parameters:
$h=0.71$, $\Omega_b h^2 = 0.022$, $\Omega_m = 0.268$, $\Omega_v = 0.732$, where
$\Omega_b$, $\Omega_m$, and $\Omega_v$ have their usual meanings as the
contributions to $\Omega$
from the gas, all matter, and the vacuum energy, respectively.  For the
hydrogen fraction by mass, $Y=0.235$.

The initial density perturbations were created by displacing a uniform
grid using the Zel'dovich approximation. The initial power spectrum of
the density fields was a {\it COBE}-normalised power law with index
$n=0.97$. The same initial conditions were used for all simulations,
except for the convergence and cosmic variance tests. Since there is no feedback from the
RT to the PM code, all the runs have identical gas densities. The
simulations were evolved to a redshift of 3.

\subsection{Cosmic variance, convergence}
\label{sec.SimVariance}

\begin{figure}
\includegraphics{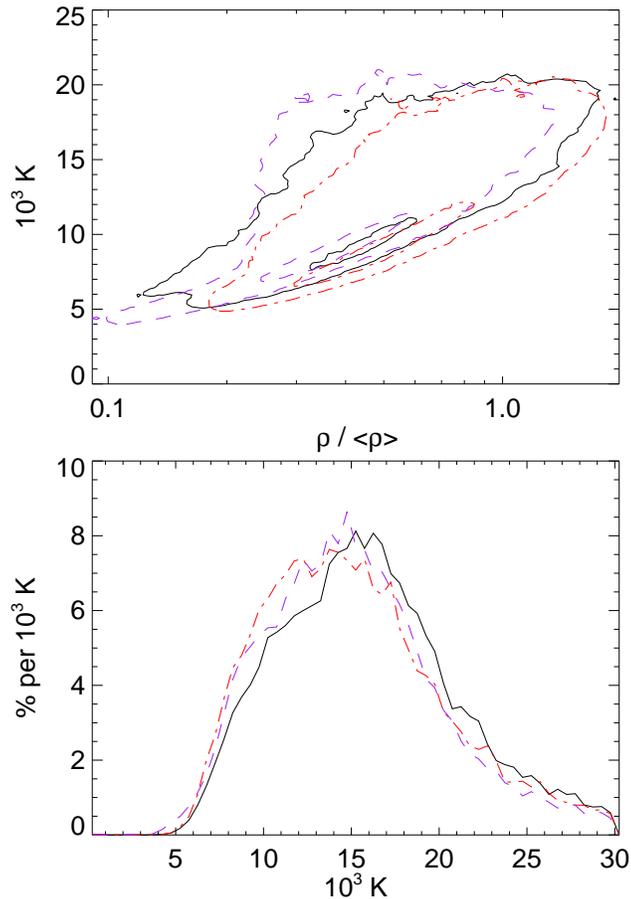}
\caption{Convergence with resolution of the $N$-body code and the
effect of cosmic variance. The top panel illustrates the $\rho - T$
distribution for the ionised gas for two $N_p = 256^3$, $N_g = 256$
realisations with the same power law model (black/solid and
red/dash-dot) and a $N_p = 512^3$, $N_g = 512$ realisation
(purple/dashed). The contours are at
$dP/d\log\left(\rho/\mean{\rho}\right)/dT =
3.6\expd{-5} \K^{-1}$ and $3.6\expd{-4} \K^{-1}$.
The bottom panel shows the mass-weighted temperature distributions for
the ionised gas for the same models.}
\label{fig.convergence_and_variance}
\end{figure}

The results are subject to cosmic variance and bias from the
resolution of the $N$-body code and the RT grid, but not
substantially. In this section we describe two extra simulations which
ascertain the effects of cosmic variance and a change in the
resolution parameters by comparing these with the PL08 run which they
most closely match. We use the $\rho-T$ and temperature distributions
(\fig{fig.convergence_and_variance} top and bottom panels) to
illustrate the effects.  The extra simulations also show that the lack of advection in the simulations is not a source of significant error.

There are two parameters that control the spatial resolution of the
simulation: the number of particles in the PM simulation, $N_p$, and
the mesh size of the gas density grid, $N_g$. To estimate the
variation due simply to cosmic variance, a simulation was run with a
different realisation of the initial density fluctuations at the same
resolution as the main body of runs ($N_p = 256^3$, $N_g = 256$). In
\fig{fig.convergence_and_variance}, the difference between the solid
(black) and dot-dashed lines (red) indicates the cosmic variance.
A simulation with 8 times the mass resolution ($N_p = 512^3$) and
twice the RT mesh resolution ($N_g = 512$) produced a qualitatively
equivalent distribution (\fig{fig.convergence_and_variance}, dashed
line), bracketed by the two lower resolution realisations.

The current code does not advect thermodynamic quantities. Hence if a dense halo has a high velocity, the gas in the halo is not properly modelled. The error from the lack of advection is not easy to quantify, but it is not believed to be large.  The error is more prominent when cells are smaller since moving gas is more likely to cross a smaller cell.  The comparison of the $256^3$ runs to the higher-resolution $512^3$ run (\fig{fig.convergence_and_variance}) demonstrates little variation.

\section{Simulation results}
\label{sec.Results}
The simulations provide information for two epochs of interest:\ the
period during which the gas is being ionised and the post-ionisation
epoch. Our primary focus in this paper is the post-ionisation
temperature of the gas, but future telescopes such as the {\it Low
Frequency Array}
\footnote{www.lofar.org} ({\it LOFAR}), the {\it Mileura Widefield Array}
\footnote{web.haystack.mit.edu/arrays/MWA/site/index.html} ({\it MWA}),
and eventually the {\it Square Kilometre Array}
\footnote{www.skatelescope.org} ({\it SKA}),
will be able to probe the epoch of ionisation itself via the red
shifted 21cm line of neutral hydrogen \citep[see][for a review]{FOB06}. The reionisation epoch will be
the topic of a future paper.

In this section, we focus on the thermal state of the gas as
established by the passage of an individual ionisation front. The
reionisation process involves the overlap of such fronts.  The
ionisation fractions of the gas after the reionisation epoch will be
constantly reset as the gas sees ionising photons from an increasing
number of sources, both because of the time needed for photons to
reach the gas and as new sources turn on (while older ones die). These
effects are not accounted for by our single I-front models. The {\it
temperature} of the gas, however, is nearly insensitive to the
continual resetting of the level and shape of the ionising photon
background \citep{Meiksin94, HG97}. It is instead
determined by the reionisation process itself and the subsequent
evolution of the physical state of the IGM. We here explore the
evolution of IGM temperature as computed in our simulations. A
principal goal is to determine if the temperature of the ionised IGM
is dependent on the nature of the ionising source. We find that it is,
and in \sect{sec.TemperatureResults} we quantify and discuss the
magnitude of this effect and some of its consequences for the
ionisation structure of the IGM.

The simulation results also permit us to explore a few other issues
related to the reionisation of the IGM. Clumping of the gas will
generally impede the propagation of ionisation fronts. We evaluate the
importance of this effect in \sect{sec.Clumping}. Because of the
inclusion of helium ionisation in our simulations, we are able to
examine the ratio of \HeII\ and \HeIII\ to \HI\ in the ionised IGM
prior to the epoch of complete \HeII\ reionisation. We discuss these
results in \sect{sec.Helium}.

But first, to justify the effort put into the implementation of RT in
our code, we make a comparison with the optically thin approximation
in \sect{sec.OTA}.

\subsection{Comparison with the optically thin approximation}
\label{sec.OTA}

Neglecting RT can underestimate temperatures by a factor of a few \citep{AH99,
BMW04, BH06}.
The importance of including radiative transfer effects during
reionisation to obtain accurate temperatures is best illustrated by
comparing against a simulation using the optically thin approximation
(OTA). Simply put, the OTA means all points not near a source receive
the same average radiation field. The approximation works fairly well
because any gas dense enough to self-shield is also dense enough to
reach thermal balance. So the post-ionisation temperature is
essentially insensitive to the details of reionisation. The
approximation fails, however, to properly treat low density gas,
because the time to reach thermal equilibrium in low density gas
exceeds the Hubble time at high redshifts. As a consequence, the low
density gas retains a memory of the reionisation details. This is
particularly important when low density gas is shielded from the
oncoming I-front by dense structures because of their role in
hardening the spectrum of the radiation field.

\begin{figure}
\includegraphics{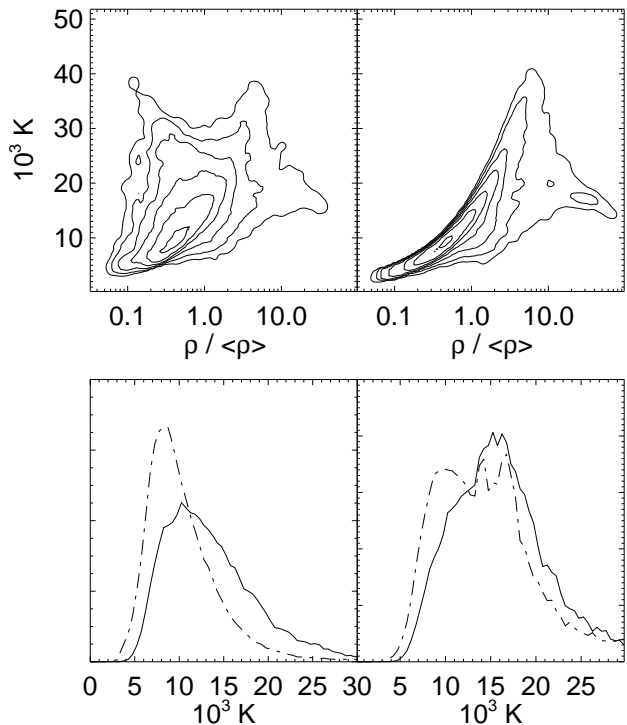}
\caption{Effects of optically thin approximation. The top panels
illustrate the $\rho-T$ distribution without (left) and using (right)
the optically thin approximation. The lower panels show the
modification to the volume-weighted (left) and mass-weighted (right)
temperature distributions produced without (solid) and with
(dot-dashed) the approximation. The volume-weighted distribution is
affected more because most of the volume of the Universe contains
underdense regions, which retain a memory of the reionisation process.}
\label{fig.Compare_OT}
\end{figure}

To compare our results with those using the OTA, we performed a
simulation in which the cumulative optical depth to any point was set
to zero, mimicking the OTA. The $\rho-T$ and temperature distributions
for the PL08 model with and without the approximation are illustrated
in \fig{fig.Compare_OT}. Overall the distributions are similar. Since
different regions are exposed to different spectra without the OTA,
there is more spread than when the OTA is used. The hardening of the
spectrum due to selective absorption of low-energy photons heats the
gas, particularly in the less dense regions. A high density spur of
decreasing temperature with increasing density is produced in both
simulations, resulting from the establishment of thermal balance
between photoionisation heating and atomic cooling \citep{Meiksin94}.

\subsection{Clumping}
\label{sec.Clumping}

\begin{figure*}
\includegraphics{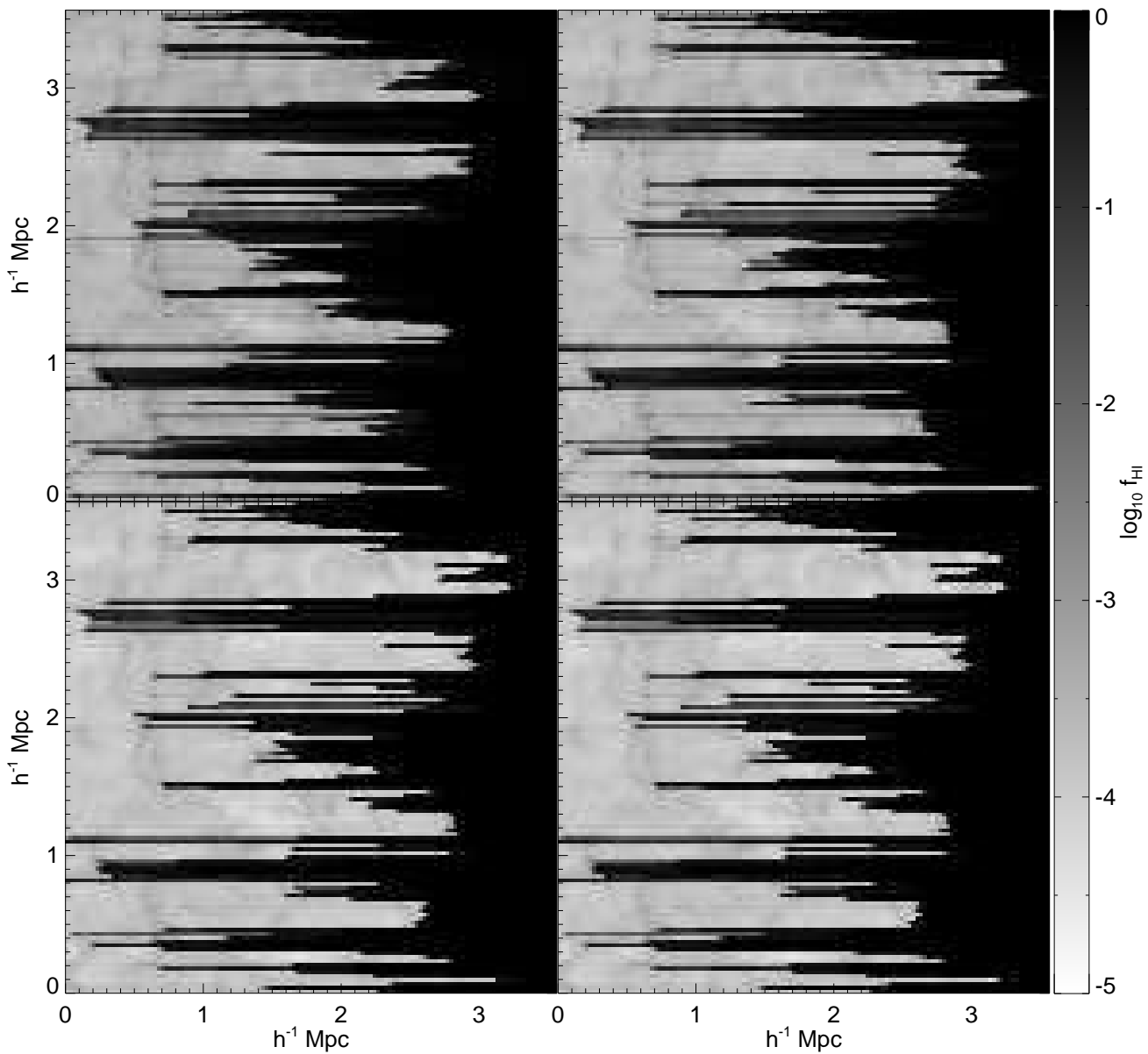}
\caption{The \HI\ fraction at $z=6$. From top-left to bottom-right
the panels correspond to PL20, MQ20, SB20, and HY20. (The hard source has
not yet turned on in the hybrid model, so the simulation at this point
is identical to the starburst run.) The ionising flux is incident from the
left. Distances are in proper units.}
\label{fig.H1_fraction_maps}
\end{figure*}

It has long been recognized that clumping of the IGM gas may
substantially slow the propagation of I-fronts due to the increased
rate of radiative recombinations \citep{SG87, MM93, MM94}. Estimates
of the importance of clumping have varied, but recently tend toward
only a moderate slowdown of the I-fronts \citep{SAH03, Meiksin05,
Ciardi06} with the resistance increasing with time as clumping increases
\citep{ISS05}. In our simulations, the slowing of the I-fronts in
individual lines of sight leads to shadows in the ionisation maps, as
shown in \Fig{fig.H1_fraction_maps}.

\begin{figure}
\includegraphics{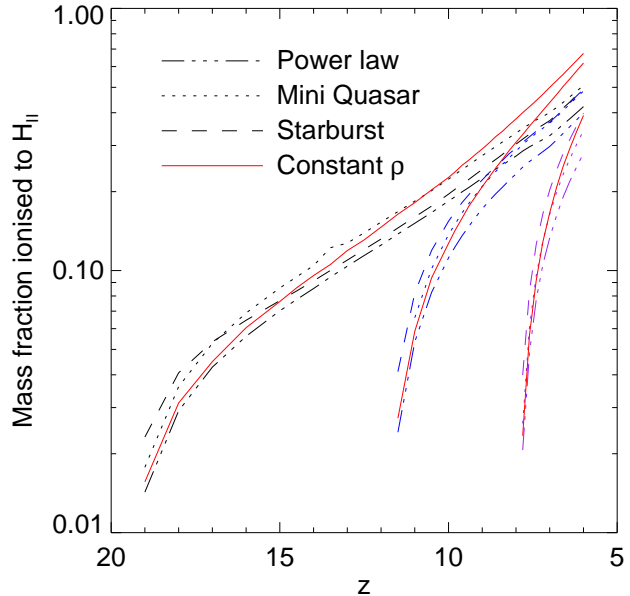}
\caption{Fraction of the simulation volume reionised for three models,
compared with the case for a uniform density IGM at the mean density.
The hybrid model follows the same history as the starburst model
prior to $z=6$. A simple model in which a uniform density gas ($\rho =
\Omega_g \rho_c [1+z]^3$) is ionised by a constant flux of ionising
photons is plotted as the solid line. There are three separate
redshifts at which the ionising flux is turned on. All the models
initially ionise the gas faster than the uniform density case as the
I-front sweeps into underdense regions. Clumping eventually slows down
the growth of the \HII\ filling factor by an amount depending on the
spectral shape.}
\label{fig.filling_factors_HII}
\end{figure}

We have estimated the role clumping may play in delaying reionisation
by comparing the growth of the \HII\ filling factor in the simulations
with that predicted for a uniform IGM at the mean baryon density up to
$z=6$, at which point overlapping I-fronts would complete the
reionisation. Introducing structure into the IGM actually results in a
moderate increase in the rate at which the filling factor of ionised
gas grows in the early stages, as shown in
\fig{fig.filling_factors_HII}. This is because most of the volume of
the Universe is underdense. Once half of the volume is reionised, the
filling factor converges to the uniform density prediction, with a
small slowdown depending on the spectral shape of the source. The
evolution of the mass-weighted fraction of ionised hydrogen follows a
similar trend including the more rapid rise than the uniform density
case at early times. By $z=6$, however, the fraction grows
substantially less rapidly, with 50 per cent to a factor of two less
mass ionised than for the uniform density prediction. The slower
growth for the mass-weighted case vs the volume filling factor is
expected since it takes longer for the I-front to penetrate the
densest regions which contain proportionally more mass.

It is possible the amount by which the I-front slows down is
underestimated because of the deficit of
small dense structures like Lyman Limit Systems in $N$-body simulations \citep{GKHW97,
MW04}. A definitive result may need to await hydrodynamical
simulations that reproduce the statistics of Lyman Limit Systems and
denser intergalactic structures.

\subsection{Temperature}
\label{sec.TemperatureResults}

\begin{figure*}
\includegraphics{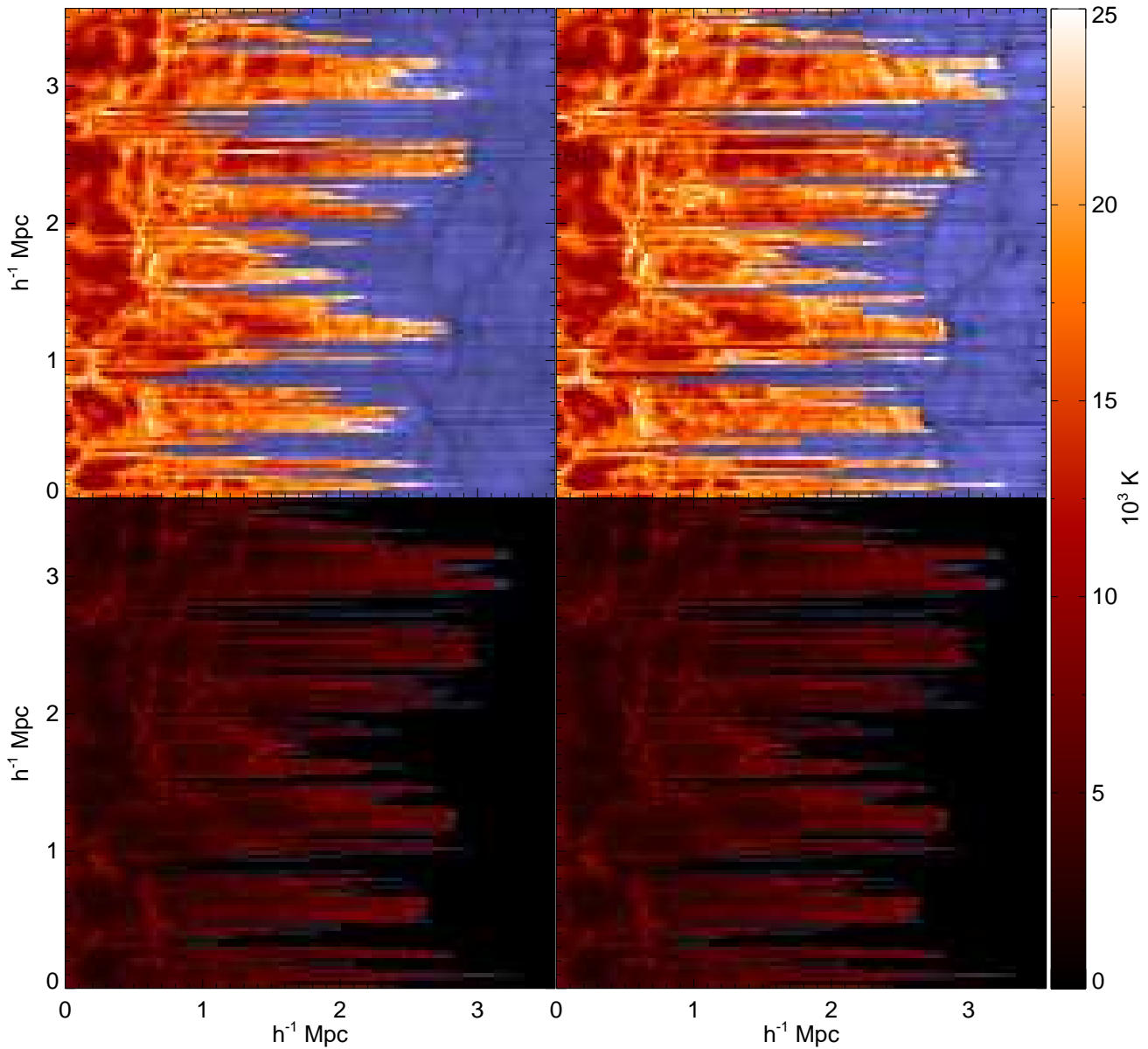}
\caption{The temperature at $z=6$. From top-left to bottom-right the
panels correspond to PL20, MQ20, SB20, and HY20. (The hard source has
not yet turned on in the hybrid model, so the simulation at this point
is identical to the starburst run.) The ionising flux is incident from the
left.
Only regions for which $f_{\rm HI}<0.1$ are shown. The remaining gas
(masked as blue) would be ionised by other sources.}
\label{fig.T_maps}
\end{figure*}

\begin{figure*}
\includegraphics{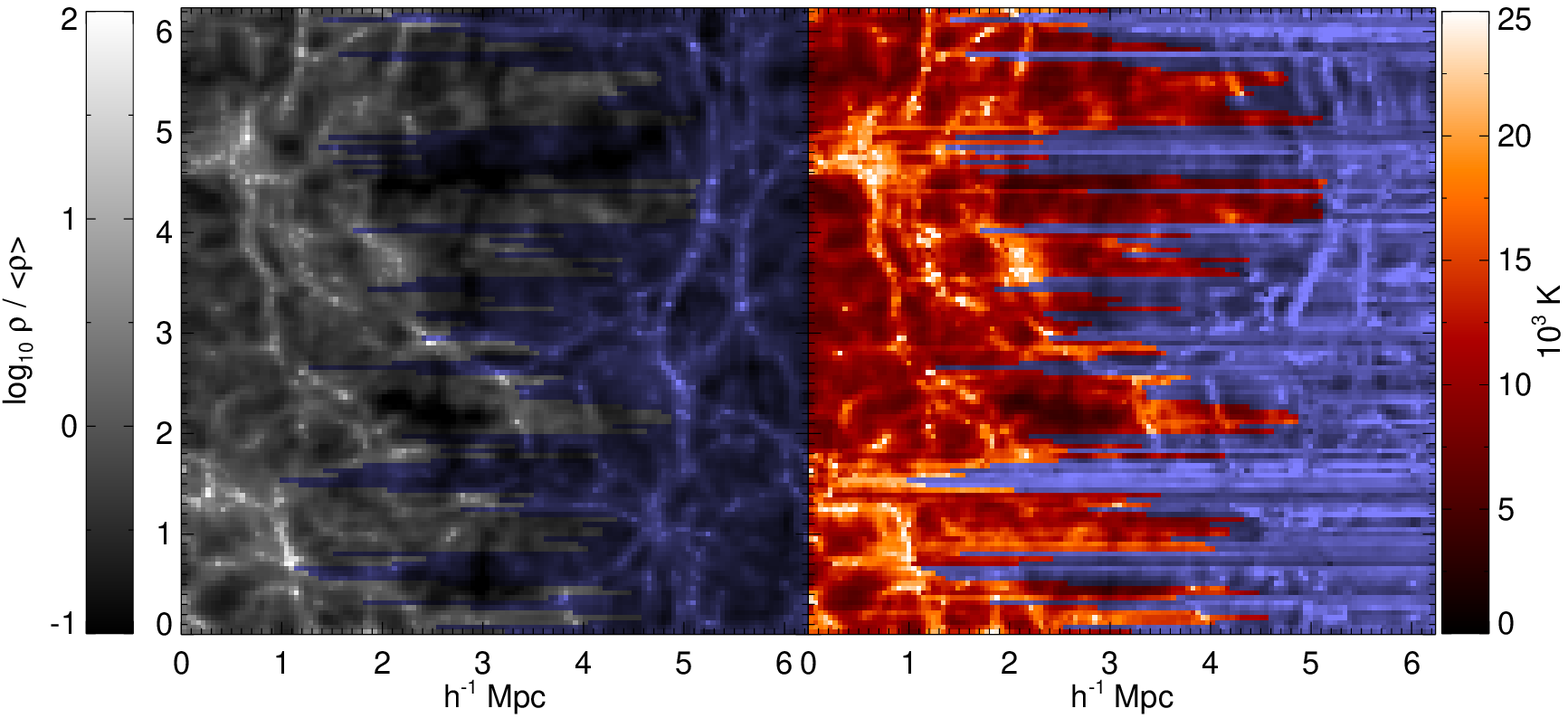}
\caption{Density distribution (left) and temperature map at $z=3$.
The temperature map is from PL20. Only regions
for which the gas is ionised prior to $z=6$ are shown, the remaining regions
are masked as blue.}
\label{fig.rho_T_maps}
\end{figure*}

\Fig{fig.T_maps} is a map of the temperature distribution at a
redshift of $z=6$. \Fig{fig.rho_T_maps} shows a similar map for the
power-law model at $z=3$. In both maps, only regions in which the \HI\ 
was ionised ($f_{\rm HI}<0.1$) at $z=6$ are shown. (The remaining gas
would have been ionised by a different ionisation front or fronts by
this time.)  We immediately see a variety of effects that will be
explored quantitatively later. First, the gas temperature in the
miniquasar model is virtually identical to that of the simple
$\alpha=0.5$ power law. Second, the starburst model produces the
coolest ionised gas. (At this stage, the hybrid and pure starburst
models are identical.) Third, the highest temperatures are typically
found just behind the \HII\ I-front. Fourth, a ``streaking'' effect is
apparent. The gas temperature downstream of a dense clump of gas is enhanced
compared with the surrounding gas as a result of delayed
ionisation.  The enhancement persists until $z=3$. Finally, when compared with the gas density, all the
models produce gas temperature structure that traces the
large-scale gas structure, as illustrated in \fig{fig.rho_T_maps}. The
relation, however, is not one of simple proportionality, as we will
see below. For instance, because the most recently ionised gas tends
to be hotter at a given density, the temperature tends to increase
towards the right (because the I-front passes from left to right), as
shown in \fig{fig.T_maps}.

\begin{figure}
\includegraphics{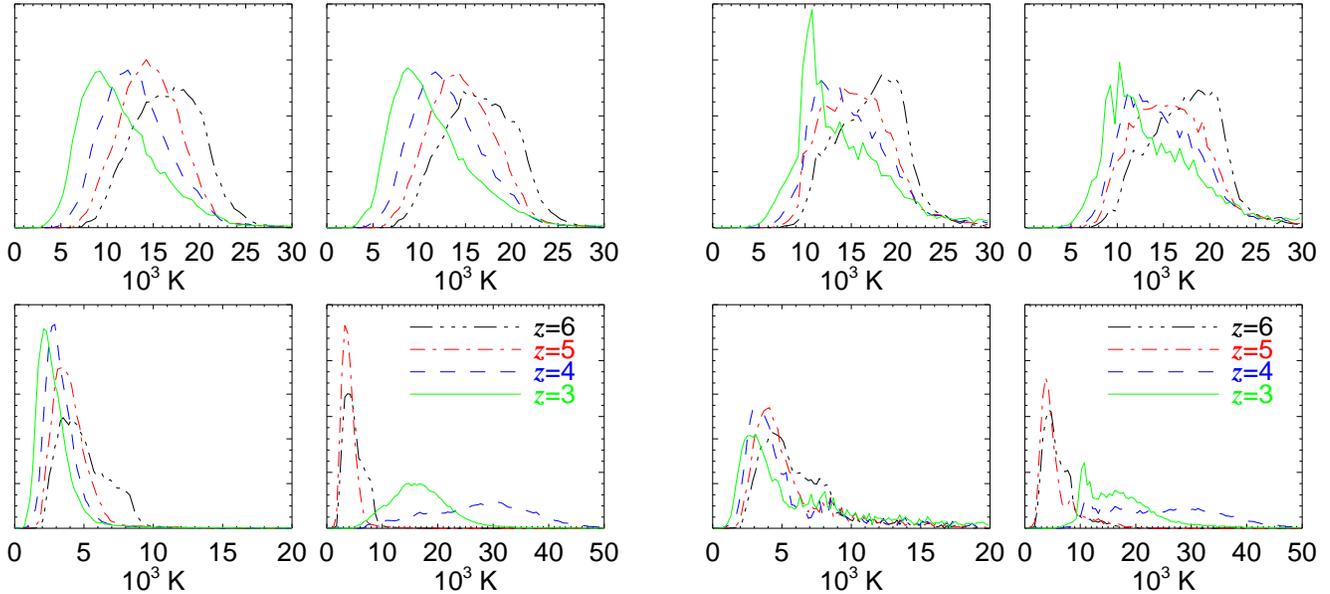}
\caption{The volume-weighted temperature distribution at $z=6$, 5, 4 and 3
for the gas ionised before $z=6$ only. From top-left to bottom-right the
panels correspond to the power-law, miniquasar, starburst, and hybrid model.
The colours/line type distinguish the redshifts, with
$z = 6$ black/dash-dot-dot-dot, $z = 5$ red/dash-dot,
$z = 4$ blue/dashed, and $z = 3$ green/solid.
Note the change in the temperature scale for the various models.}
\label{fig.T_dist_ionised}
\end{figure}

\begin{figure}
\includegraphics{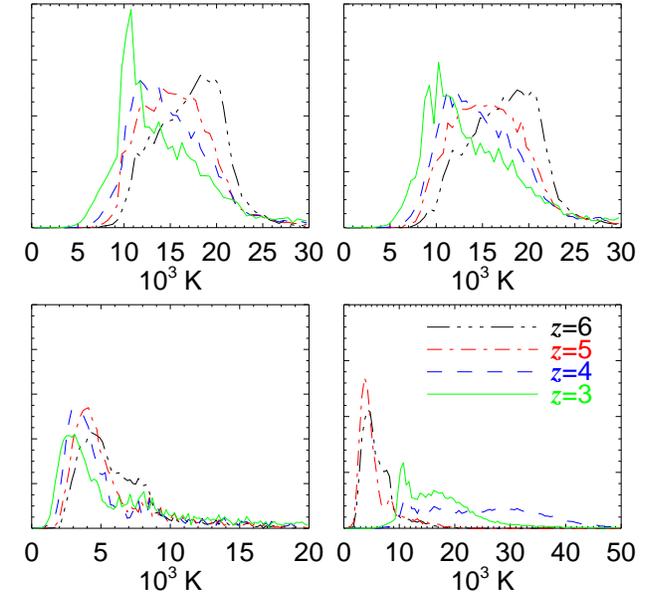}
\caption{The mass-weighted temperature distribution at $z=6$, 5, 4 and 3
for the gas ionised before $z=6$ only. From top-left to bottom-right the
panels correspond to the power-law, miniquasar, starburst, and hybrid model.
The colours/line type distinguish the redshift, with
$z = 6$ black/dash-dot-dot-dot, $z = 5$ red/dash-dot,
$z = 4$ blue/dashed, and $z = 3$ green/solid.
Note the change in the temperature scale for the various models.}
\label{fig.T_dist_mass_weighted_ionised}
\end{figure}

The gas temperatures distinguish the various models at all redshifts. The
temperature distributions of the gas ionised by $z=6$, both volume-weighted
(\fig{fig.T_dist_ionised}) and mass-weighted
(\fig{fig.T_dist_mass_weighted_ionised}), show clear differences for
the different model spectra. At $z=3$, for the power-law and
miniquasar models, the temperatures span 8 to
$28\expd{3}\K$ (90 per cent of the gas mass with 5 per cent below the range
and 5 per cent above) with a mass-weighted mean of $15\expd{3}\K$. The hybrid model
has a hotter tail in its distribution, ranging from $\sim 10$ to
$31\expd{3}\K$ with a mass-weighted mean of $17\expd{3}\K$.  Gas two to three times cooler
is produced by the starburst model. For this model, the temperatures
at $z=3$ range from 2000 to 17000~K with the mass-weighted mean at $9000\K$. The
distributions are, however, highly skewed with mass-weighted modes at $11000\K$ for the power-law, miniquasar, and hybrid spectra and $3000\K$ for the starburst spectrum.

There are few direct determinations of the IGM gas temperature.  The measured Doppler parameters on their own provide only an upper limit since the lines may be velocity broadened (e.g. due to microturbulence). If metal lines are present, however, the thermal and kinematic contributions are separable.  
By simultaneously fitting \CIV\ and \SiIV\ absorbers, \citet{RSWB96} found average temperatures of $\simeq 38000\K$ for systems with $\log_{10} n_\HIs > 14$. Only the hybrid model is able to achieve temperatures of $T\simeq 40000\K$ (\figs{fig.T_dist_ionised} and \ref{fig.T_dist_mass_weighted_ionised}).

We have tested that the redshift at which the source turns on is not a
significant factor.
Save for a slight shift to higher temperatures for the $z_{\rm
on}=8$ models, the curves are nearly identical. This is because the incident
ionising flux has been normalised to a common value for all the cases so
that by $z\lsim8$ the reionisation proceeds in a nearly identical manner.
Changing the epoch of reionisation would change the temperature in the low
density gas at later times. This effect is not explored here.

\begin{figure}
\includegraphics{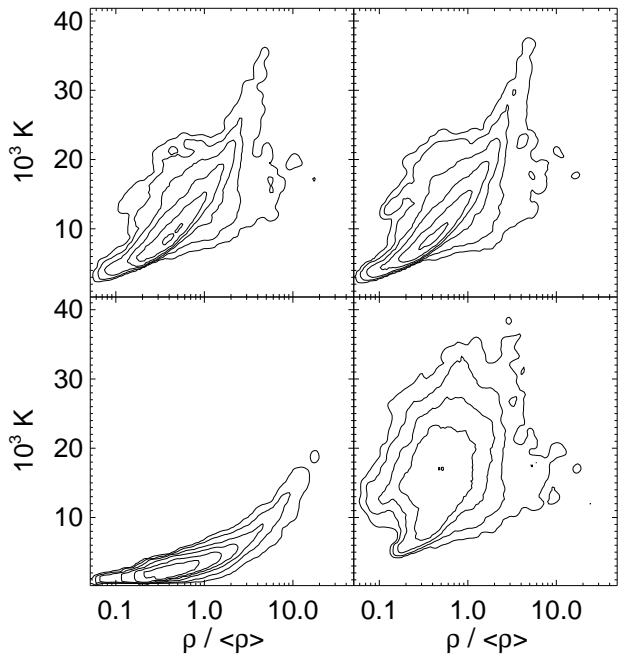}
\caption{The $\rho-T$ distribution at $z=3$ for the gas ionised by $z=6$ only.
From top-left to bottom-right the panels correspond to PL20, MQ20,
SB20, and HY20. The contour intervals are spaced by a factor of $\sqrt
10$.}
\label{fig.rho_T_dist_ionised}
\end{figure}

As we showed above, the use of full radiative transfer instead of an
optically thin approximation not only further heats the gas, but it
spreads the $\rho-T$ distribution away from a tight power law
(\fig{fig.Compare_OT}). The spread in temperatures is greatest for gas
with $\rho/\bar{\rho} < 1$. The spectrum of the ionising radiation is
modified by preferential absorption of the lower frequency photons as
they pass through the gas. The modification hardens any spectrum, but
is most influential to spectra that already have a hard
component. Hence, the spread should be largest for harder spectra.
\Fig{fig.rho_T_dist_ionised}, which maps the $\rho-T$ distribution at
$z=3$ for the ionised gas for PL20, MQ20, SB20, and HY20, confirms the
larger spread in the harder spectra. Also confirmed is the similarity
of the state of the gas in the models with the power-law and
miniquasar spectra.

\begin{table*}
\caption{Fits to the polytropic relation, $T=T_0(\rho/\bar{\rho})^{\gamma-1}$
of the gas for which $f_\HIs < 0.1$ by $z=6$.
The errors are to the fits and not indicative of the distribution in the
$\rho-T$ plane.}
\begin{tabular}{|c|c|c|c|c|c|}
\multicolumn{2}{c}{} & \multicolumn{4}{c|}{$z$} \\\cline{3-6}
\multicolumn{2}{c}{\raisebox{1.5ex}[0cm][0cm]{ }}
 & 6 & 5 & 4 & 3 \\\hline
 & $T_0 [\K]$ & $17900\pm 100$ & $16900\pm 150$ & $15400\pm 140$ & $14600\pm 200$ \\
\cline{2-6}
\raisebox{1.5ex}[0cm][0cm]{PL20}
 & $\gamma$ & $1.29\pm 0.01$ & $1.40\pm 0.01$ & $1.52\pm 0.01$ & $1.58\pm 0.01$ \\
\hline
 & $T_0 [\K]$ & $17900\pm 120$ & $16950\pm 170$ & $15500\pm 150$ & $14500\pm 160$ \\
\cline{2-6}
\raisebox{1.5ex}[0cm][0cm]{MQ20}
 & $\gamma$ & $1.34\pm0.01$ & $1.44\pm 0.01$ & $1.52\pm 0.01$ & $1.57\pm 0.01$ \\
\hline
 & $T_0 [\K]$ & $4200\pm 130$ & $4070\pm 50$ & $3720\pm 20$ & $3515\pm 11$ \\
\cline{2-6}
\raisebox{1.5ex}[0cm][0cm]{SB20}
 & $\gamma$ & $1.44\pm 0.03$ & $1.43\pm 0.01$ & $1.47\pm 0.01$ & $1.553\pm 0.002$ \\
\hline
 & $T_0 [\K]$ & $4200\pm 110$ & $4040\pm 50$ & $29250\pm 250$ & $18600\pm 100$ \\
\cline{2-6}
\raisebox{1.5ex}[0cm][0cm]{HY20}
 & $\gamma$ & $1.42\pm 0.03$ & $1.42\pm 0.01$ & $-0.0363\pm 0.009$ & $1.134\pm 0.005$ \\
\hline
\label{table.PolyFits}
\end{tabular}
\end{table*}

Pseudo-hydrodynamical models of the IGM usually adopt a polytropic
equation of state for the gas. We fit a polytropic relation, $T = T_0
(\rho/\bar{\rho})^{\gamma-1}$, to the $\rho-T$ distributions at $z=6$,
5, 4 and 3. The results are shown in \tab{table.PolyFits}.
Note that the errors are for the coefficients and not an indication of
the spread of the $\rho-T$ distribution, which is not particularly
well-described as a single polytropic dependence. For comparison, for
the OTA run, for which local heating is balanced by local cooling, we
find $T_0 = 13780\pm 40$ and $\gamma = 1.527\pm 0.003$ at $z=3$, with
less spread than found when radiative transfer is
incorporated. Together, these very different simulations imply a local
balance between heating and cooling still dominates the setting of the
polytropic index.

Generally $\gamma=1$ is assumed to be the theoretical lower limit,
corresponding to isothermality. Most of the fits indeed show
$\gamma>1$. The exception is the value of $\gamma$ for the hybrid
model. Soon after the QSO starts ionising \HeII, lower values of
$\gamma$ occur, including $\gamma<0$ at $z=4$. This arises because of
the poor representation a polytrope gives of the wide spread in
temperatures. Any applications restricting to $\gamma>1$ may
therefore be excluding a description of the thermal behaviour of the
IGM during the \HeII\ reionisation epoch. At the other extreme, for
all models the index is lower than the adiabatic index, $\gamma_{\rm
ad}=5/3$, reflecting the influence of radiative cooling in
high-density regions.

The polytropic fits resulting from the simulations may be compared
with those of \citet{Schaye00} to Keck HIRES spectra. These authors
find $T_0\approx2.2\pm0.2\times10^4$~K and $\gamma\approx1.0\pm0.1$ at
$z\approx3$, values most consistent with the late \HeII\ reionisation
hybrid model.

\subsection{Ionisation rates}
\label{sec.IonisationRates}

\begin{figure}
\includegraphics{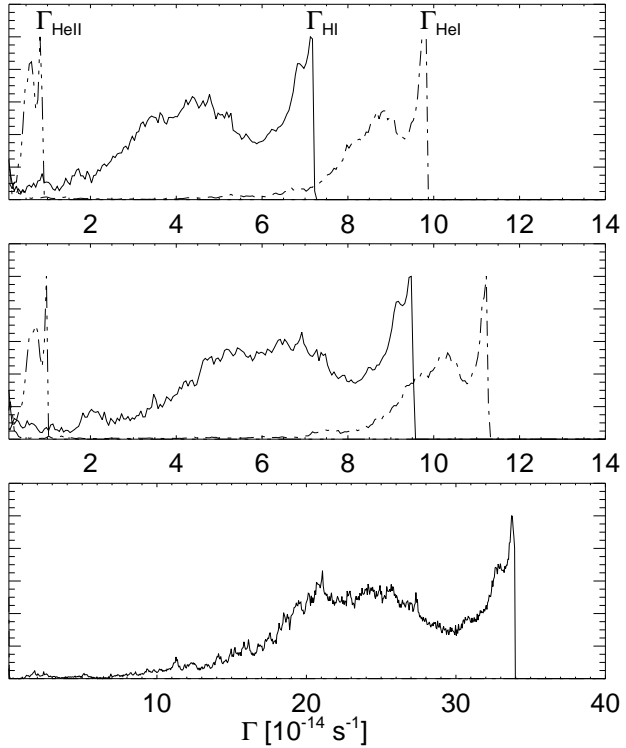}
\caption{Distribution functions for the ionisation rates at $z=6$.
The distributions from the various species are distinguished as
indicated. Only rates from regions with ionised hydrogen ($f_{\rm
HeII}<0.1$) are shown. From top to bottom, the panels correspond to
the power-law, miniquasar, and starburst models. (The hybrid model is
identical to the starburst model at this epoch.)  The only
contribution to the starburst distribution comes from hydrogen. Note
the different $\Gamma$ range for the starburst model. Also note that
the distributions for the individual species are normalised
independently for clarity. All the distributions show extensive
tails toward low values resulting from the absorption of photoionising
photons by the IGM.}
\label{fig.Gamma_dist}
\end{figure}

The varying optical depth to ionising photons results in fluctuations
in the ionising background. The ionisation background is parametrized
by the ionisation rate, $\Gamma$, which is the number of
photoionisations per atom per unit time. A long tail towards low
ionisation rates is found, as shown in \fig{fig.Gamma_dist}. The sharp
cut-off at the high end corresponds to low optical depth to the source
leading to negligible filtering of the source spectrum. Filtering decreases the ionising photon flux, lowering $\Gamma$.

A tail in the $\Gamma$-distribution is similarly found in
the simulations of \citet{MF05}, although without the sharp cut-off at
the high end. They model a uniform source distribution filtered
through the IGM, accounting for the radiative transfer of the ionising
photons using a Monte Carlo algorithm.  The absence of the sharp upper
cut-off in the simulations of \citet{MF05} is consistent with the
discreteness effects of their Monte Carlo radiative transfer
approach. In the presence of a population of randomly
distributed local ionisation sources, \citet{MW03} show that the sharp
cut-off would be replaced by a power-law tail varying as $\Gamma_{\rm
HI}^{-2.5}$.

\subsection{Helium}
\label{sec.Helium}
The presence of helium alters the temperature of the IGM after reionisation
by an amount that depends on the reionisation scenario. The epoch of helium
reionisation (ionisation of \HeII\ to \HeIII) is still unknown. Measurements
of the \HeII\ \Lya\ optical depth suggests it occurred at $z\lsim3$
\citep{Zheng04,Reimers05}, which is consistent with the expected epoch of
\HeII\ reionisation by QSO sources with soft spectra \citep{Meiksin05}. 
Measurements of the \Lya -forest Doppler parameter, $b$, permit an estimate of the IGM temperature; \citet{Theuns02} and \citet{RGS00} claim a temperature jump of about a factor of two to three at $z \ga 3$.
During the reionisation process, and prior to its completion, large
fluctuations in the \HeII\ to \HI\ absorption signatures may be expected,
as the \HeII-ionising metagalactic UV background will show large spatial
variations.

We examine the predictions for these fluctuations
from our simulations prior to the completion of \HeII\ reionisation, by which time the \HeIII\ I-fronts have completely overlapped.

\subsubsection{Thermal effects of helium}
\begin{figure}
\includegraphics{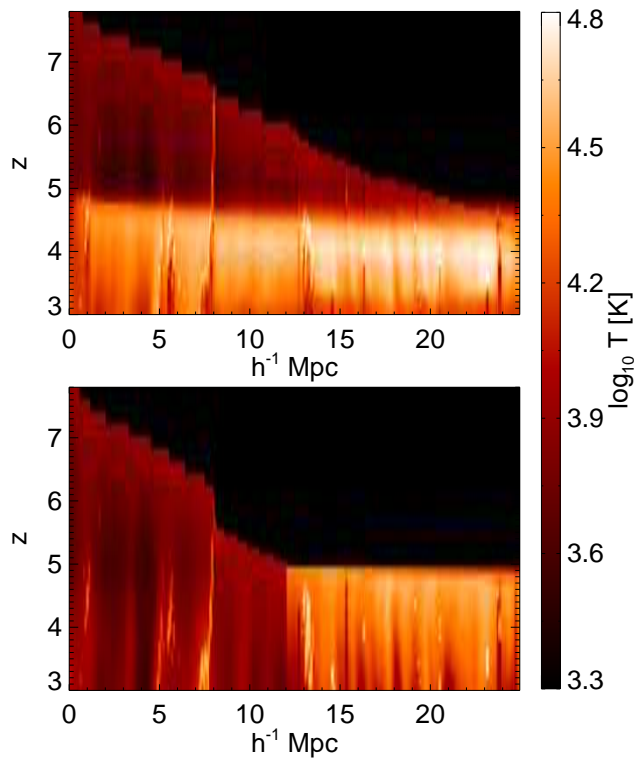}
\caption{The importance of helium in the hybrid model. Illustrated is
the evolution of the temperature of a single line of sight in a HY08
simulation with (top) and without (bottom) helium. The transition
between starburst and power-law spectrum occurs between $z=5$ and 4.}
\label{fig.HY_He_vs_no_He}
\end{figure}
The highest temperatures are produced by the hybrid model. The high
temperatures are directly attributable to the presence of helium. In
the hybrid model, the irradiating spectrum undergoes a transition from
a starburst to a power-law spectrum between $z=5$ and 4. As a
consequence, the hydrogen I-front precedes the helium I-front. Since
the temperature of ionised gas cannot be raised significantly by
changing the radiation field, no matter how hard the spectrum (barring
extreme cases like a pure x-ray spectrum), without helium the harder
power-law spectrum has no effect on the temperature of the gas
previously ionised by the starburst spectrum. Gas that has not yet
been ionised is heated to higher temperatures than that ionised by the
starburst spectrum since more energy, $h(\nu - \nu_0)$, is liberated
per ionisation. In \fig{fig.HY_He_vs_no_He}, bottom panel, the
differential heating in a hybrid model without helium ($Y=0$) is
illustrated. The gas ionised prior to the transition to the power-law
spectrum remains unaffected by the transition. The gas previously
unionised is ionised and heated to higher temperatures. With helium
(top panel of \fig{fig.HY_He_vs_no_He}) the transition of the spectrum
heats all the gas along the line of sight. This is not surprising
since the ionisation of helium introduces additional heating. What is
notable is that the gas is heated to higher temperatures than for the
case of a power-law spectrum ionising a completely neutral medium ($\sim 30000\K$ versus $\sim 16000\K$, see \fig{fig.T_dist_ionised}).

A \HeIII\ I-front passing through \HII-dominated gas leads to a larger
jump in temperature than a \HII\ front passing through \HeIII-dominated
gas. (Note that in both cases, the final state is fully ionised.) The
temperature-entropy relation, \eq{eq:TS}, accounts for the
differential temperature jumps. In the case of the \HeII-\HeIII\
transition in a \HII-dominated gas, the mean molecular weight $\mu$ is
reduced by only 7 per cent, leaving any gain in entropy to translate
into a gain in temperature. In the alternate case of the \HI-\HII\
transition in a gas where \HeIII\ is the dominate helium species,
$\mu$ drops by 45 per cent, almost halving the gain in entropy. There
is still a net gain in temperature, but it is only about 10 per cent.

The shallow $\rho-T$ profile in the hybrid model results from hotter
low-density gas compared with the other models
(\fig{fig.rho_T_dist_ionised}, lower-right panel). The high temperatures
following the recent \HeII\ reionisation are retained in the low-density
gas due to the long time for thermal equilibrium to be established in
underdense gas \citep{Meiksin94}.

\subsubsection{Correction to the \HI\ abundances at $z<6$}
Prior to overlap of the \HII\ I-fronts at $z\sim 6$, the ionisation
state of hydrogen is properly modelled by a single source, as we have
in our simulation. Equivalently, a single source is sufficient for
modelling the helium fractions prior to overlap of the \HeII\ I-fronts
at $z\sim 3$. However, at $z\la 6$, a single source is insufficient to
fix the hydrogen fractions, since once the Universe is reionised a
given region becomes exposed to a large number of ionising sources.
In particular, low-density regions once shadowed by Lyman-limit
systems along the line of sight to the source driving an advancing
I-front would be swept over by other I-fronts after the reionisation
epoch.

In order to compare the helium ionisation fractions with the hydrogen,
we need to remove the effect of the Lyman-edge shadows on the \HI\
from the simulation data. We do so by correcting the \HI\ abundances
by setting $\Gamma_\HIs$ to a uniform value at any given redshift and
solving \eq{eq.IonisationStates} for $n_\HIs$ in the static case.
Estimates of the expected distribution of $\Gamma_\HIs$ after the
reionisation epoch suggest it is narrowly peaked, especially for $z<4$
\citep{MW03}. We set $\Gamma_\HIs$ to the mean value for the
unshadowed gas at the given redshift.\footnote{This choice is
arbitrary. We could have assigned the expected value at the
corresponding redshift, as determined by matching to the measured \Lya\
effective optical depth \citep{MW04}, but since $\eta$ is proportional
to the ratio $\Gamma_{\rm HI}/\Gamma_{\rm HeII}$, it is anyway fixed
only up to an overall rescaling factor.} A negligible change to the
temperatures will be generated by an increased ionisation rate after
other sources are revealed, so that the local temperature produced
during the reionisation phase will still apply and so sets the value
of the radiative recombination coefficient $\alpha_\HIIs$ in
\eq{eq.IonisationStates}.

The solution to $n_\HIs$ is still not entirely correct, since the
dense Lyman-limit systems should contain self-shielded regions within
which the correction eliminates. The filling factor of the missing
self-shielded regions in the corrected version, however, is much less
than that of the shadowed low-density regions. We therefore applied
the correction to all the \HI\ data at $z<6$.

\subsubsection{Comparison of \HeII\ vs \HI\ absorption prior to complete helium reionisation}
By having two ionisation states, both at higher energies than that of
\HI, helium provides a means of obtaining information about the
abundance of high-energy photons. The ratio of the hydrogen and helium
column densities is a function of the spectral shape of
the ionising radiation. Coinciding \HI\ and \HeII\ absorption features
in QSO spectra have been compared by \citet{Zheng04} and
\citet{Reimers05}, who find large fluctuations in the ratio of \HeII\
to \HI\ column densities at $z\lsim3$. The presence of fluctuations places constraints on helium reionisation models.
\citet{GNBO05} find the patchiness requires short QSO lifetimes ($< 10 \Ma$) in models which attribute the patchiness to the discreteness of the QSO spatial distribution.
\citet{BHVC06} have argued for a model in which the relative number of rare \HeII -ionising QSOs compared with abundant \HeI -ionising star-forming galaxies sets the fluctuation distribution.

\begin{figure*}
\includegraphics{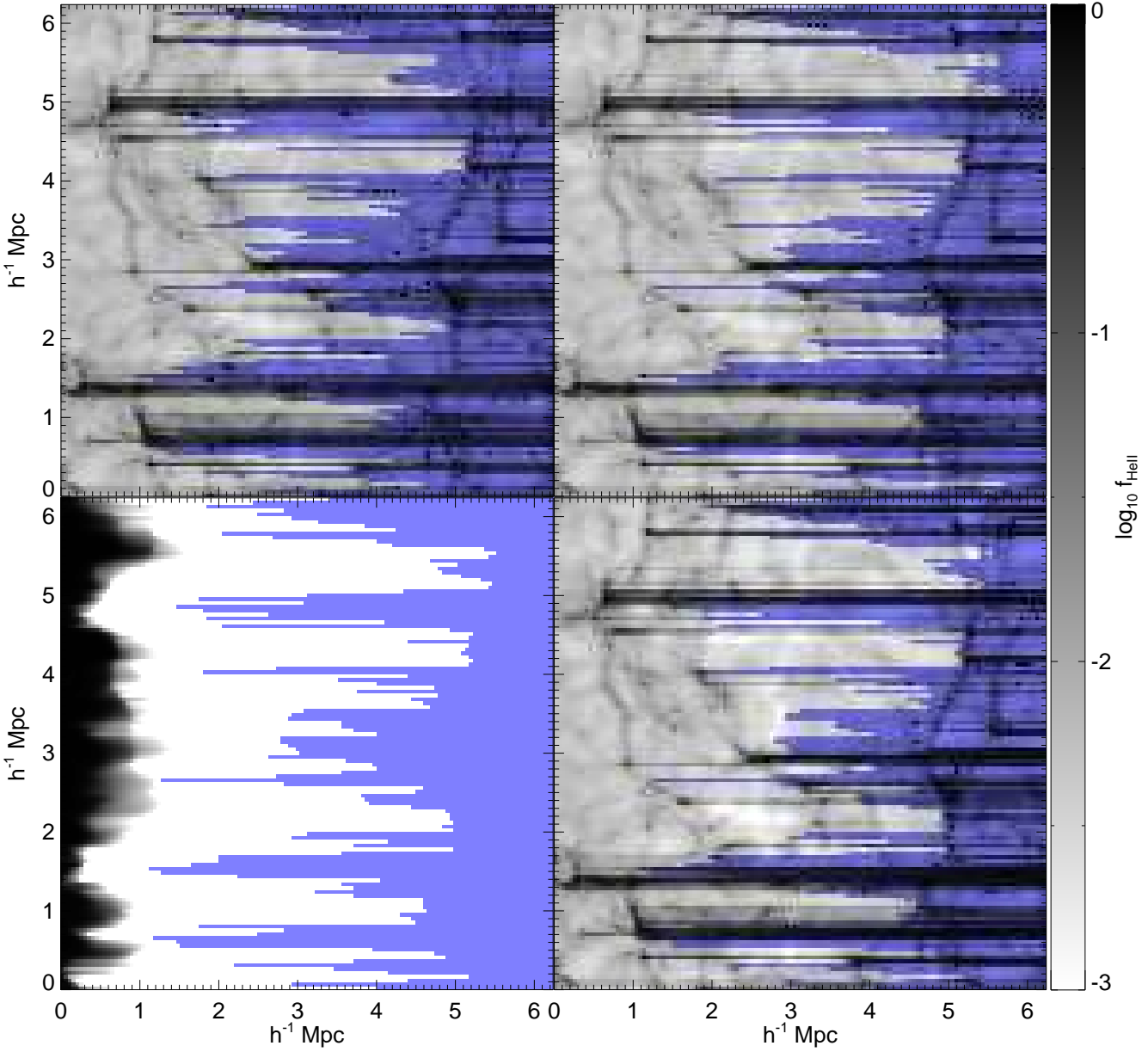}
\caption{The \HeII\ fraction at $z=3$. From top-left to bottom-right
the panels correspond to PL20, MQ20, SB20, and HY20. The regions in which hydrogen is not ionised by $z=6$ are shaded blue.}
\label{fig.He2_fraction_maps}
\end{figure*}

We confine our estimates to the amount of fluctuations expected behind
\HeIII\ I-fronts prior to their complete overlap, i.e., before the
epoch of helium reionisation has completed. We base the analysis on
the simulation results at $z=3$ (\fig{fig.He2_fraction_maps}),
which is when observations suggest \HeII\ reionisation was nearing
completion. Local sources of \HI-ionising
photons could introduce further fluctuations than those we find. Our
simulations thus only set a lower limit to the level of fluctuations
expected, arising principally from shadowing and attendant
fluctuations in the \HeII\ ionisation rate, during the final stages of
\HeII\ reionisation.

The hardness index $\eta \equiv n_\HeIIs/n_\HIs$ is a sensitive probe to the shape of the ionising background. Ionisation fraction fluctuations affect the spread in $\eta$ values while the mean value is set by the mean spectral hardness. Here we
concentrate on the relative spread of the fluctuations in $\eta$, as
this is independent of the shape of the spectrum. We do refer to
definite values for $\eta$ from our simulations rather than
arbitrarily rescaling them, but the physical effects we shall describe
are not specific to any specific value $\eta$.

\begin{figure}
\includegraphics{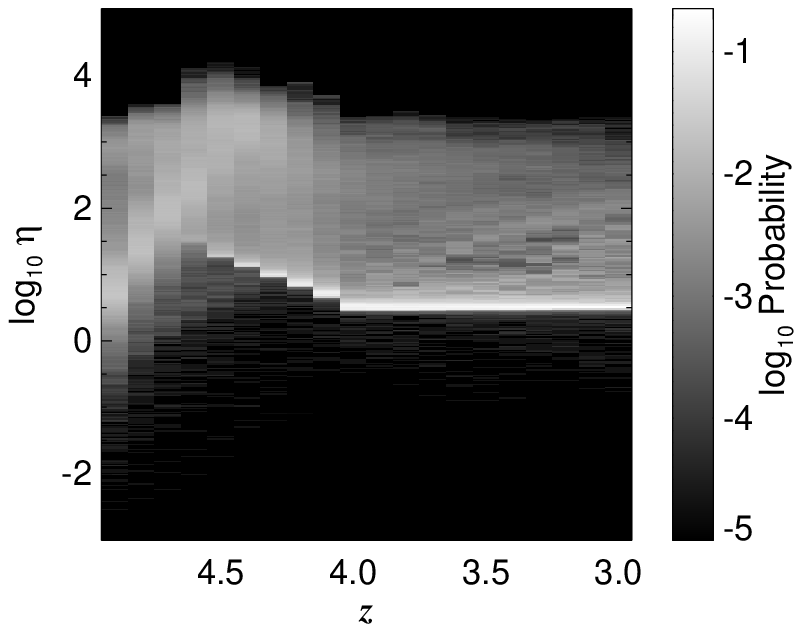}
\caption{The evolution of the distribution of $\eta=n_\HeIIs/n_\HIs$
in the hybrid model HY20, for regions in which hydrogen was reionised
by $z=6$. Although the values of $\eta$ would be
shifted by a change in the spectral shape of the ionising radiation
background, the relative spread is invariant. The source spectrum makes a transition from a starburst spectrum to a power law between $z=5$ and $z=4$. Because of the lack of \HeII -ionising photons, so that helium is not ionised to \HeIII\, the
distribution of $\eta$ is much wider at $z>5$ (not shown).}
\label{fig.eta_dist}
\end{figure}

The ionisation states of hydrogen and helium are expected to be
linearly correlated, but dependent on the hardness of the local
radiation field. This follows from radiative equilibrium, for which
the rate equations for hydrogen and helium are derivable by setting
$\dot{n}_i=0$ in \eq{eq.IonisationStates}:
\begin{eqnarray}
n_\HIs \Gamma_\HIs &=& n_e n_\HIIs \alpha_\HIIs \\
n_\HeIIs \Gamma_\HeIIs &=& n_e n_\HeIIIs \alpha_\HeIIIs . \nonumber
\end{eqnarray}
For the highly ionised gas in the simulations, $f_\HIs
\Gamma_\HIs \simeq n_e \alpha_\HIIs$ and $f_\HeIIs \Gamma_\HeIIs
\simeq n_e \alpha_\HeIIIs$, giving
\begin{equation}
\frac{f_\HeIIs}{f_\HIs} \simeq \frac{\alpha_\HeIIIs}{\alpha_\HIIs} \frac{\Gamma_\HIs}{\Gamma_\HeIIs}.
\label{eq.fractions}
\end{equation}
The ratio of $f_\HeIIs$ to $f_\HIs$ is related to $\eta$ through
$\eta=(n_\Hes/n_\Hs)(f_\HeIIs/f_\HIs)$, where $n_\Hes/n_\Hs \approx
1/13$ ($Y=0.235$) is the number ratio of helium to hydrogen atoms. 
Since \HeII\ is a hydrogen-like species, the recombination
coefficients scale similarly with temperature. Over the range 10,000 K
to 20,000 K, $\alpha_\HeIIIs \simeq 5.3\alpha_\HIIs$. Similarly, the
photoionisation rates also scale but in a more complicated manner
dependent on the spectrum (\eq{eq.Gamma}). The cross sections for \HI\ 
and \HeII\ can be approximated by using \eq{eq.CrossSection} with
$\beta$ and $s$ the same for hydrogen-like species, $\sigma_\HeIIs
= \sigma_\HIs/4$, and $\nu_\HeIIs = 4\nu_\HIs$
(\tab{table.CrossSection}). Taking the radiation field to have the
form $J_\nu \propto \nu^{-\alpha}$, the integral in \eq{eq.Gamma}
gives for the ratio of photoionisation rates $\Gamma_\HIs /
\Gamma_\HeIIs = 2^{2\alpha+2}$. Combining this result with
\eq{eq.fractions} and $\alpha_\HeIIIs \simeq 5.3\alpha_\HIIs$ gives
$\eta \simeq 5.3/13\times 2^{2\alpha+2}$. 

The values of $\eta$ found in the simulations fluctuate about this estimate. For PL20, for
example, the volume-averaged $\expectation{\eta}_{vol} = 11$ at
$z=3$.  However, the distribution is highly skewed; the mode and 68 percentile range is $3.3^{+3.5}_{-0.1}$. The derived relation predicts
$\eta \simeq 3.2$ for $\alpha=0.5$, in good agreement with the mode.  For MQ20, $\expectation{\eta}_{vol} = 14$ and, like PL20, the distribution is highly skewed with the mode $3.8^{+4.0}_{-0.1}$. The $\eta$ distribution is insensitive to redshift for the power-law and miniquasar models. In the case of the hybrid model, the evolution of the $\eta$ distribution is complex after the power-law source turns on.  For HY20, the evolution of the $\eta$ distribution is illustrated in \fig{fig.eta_dist}.  As the \HeIII\ I-front sweeps through the volume, $\eta$ surges to $>1000$. At the same time, the competing effect of the hardening spectrum drives the most probable $\eta$ values down.  After the source spectrum has completed the transition to a power law, the most probably $\eta$ value settles at the expected value of $\eta \simeq 3.2$ while the shadowed regions provide a high-$\eta$ tail.

\begin{figure}
\includegraphics{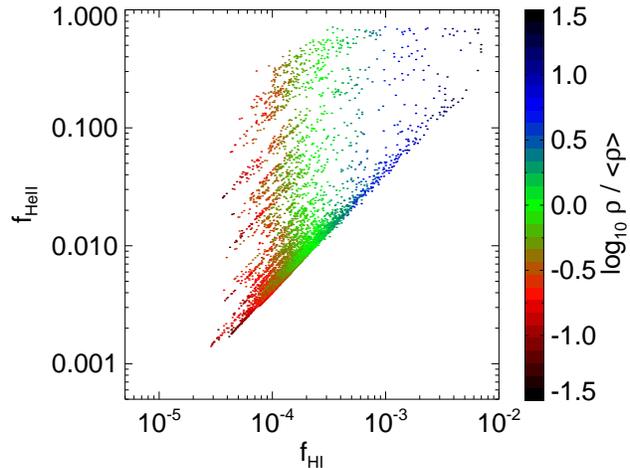}
\caption{\HeII\ versus \HI\ fractions showing dependency on density at
$z=3$ for model PL20, but with a uniform \HI\ photoionisation rate (see text). The 
MQ20 and HY20 results are qualitatively similar while those for SB20 are uncorrelated owing to the paucity of helium-ionising photons. Only regions for which $f_{\rm HI}<0.1$ by $z=6$
are shown.}
\label{fig.f_h1_vs_f_he2}
\end{figure}

The fluctuations that produce the high-end tail of the $\eta$ distribution at $z=3$ arise from the attenuation of the source radiation field at the \HeII\ ionisation threshold. Recall that we have corrected the data to remove $f_\HIs$ fluctuations, hence softer (larger values of $\alpha$) local ionising fields are generated by attenuation. Variations in the local density have only a small role as they modify the local gas temperature which varies the approximation $\alpha_\HeIIIs \simeq 5.3\alpha_\HIIs$. \Fig{fig.f_h1_vs_f_he2} illustrates the interplay of effects.  The bulk of the gas resides along the $f_\HeIIs / f_\HIs \simeq 5.3\times 2^{2\alpha+2}$ locus in the $f_\HeIIs-f_\HIs$ plane.  Attenuation leads to shadows with larger $f_\HeIIs$ fractions, creating parallel loci of constant $f_\HeIIs / f_\HIs$. The local overdensity sets the value of $f_\HIs$, as we have set $\Gamma_\HIs$ to a constant value. Because of the correction to $n_\HIs$, the effects of self-shielding are not seen in the $\eta$ distribution. Self-shielding would harden the spectrum, generating $\eta$ values below the bulk of the gas. This effect is seen in the uncorrected simulation data in the few regions in which it occurs.

The magnitude of $\eta$ is sensitive to the input source spectra. For
instance, choosing $\alpha=2$ instead of 0.5 for the power-law
spectrum would increase $\eta$ by a factor of 8 for the power-law
model to values of $\eta\approx300$. Boosting the ratio of the
contribution of the galaxy to the QSO spectrum in the hybrid model
would achieve a similar effect. The relative spread in the
fluctuations of $\eta$, however, is independent of any overall shift
in the amplitude or shape of the ionising background for regions in
which both hydrogen and helium are ionised. The spread may be a useful
diagnostic of the ionisation state of the IGM prior to complete \HeII\
reionisation. Large fluctuations are found for $\eta$ (or,
equivalently, in $f_{\rm HeII}/f_{\rm HI}$). At $z=3$, $\eta$ is
mostly constant, but there is small fraction with a large range ($\sim
2$ dex). The extent of the range results both from the inhomogeneities
in the radiation field and a wide spread in gas temperatures due to
radiative transfer, particularly in the low density gas which gives
rise to most of the \HeII\ features. The spread is smaller than found
by \citet{Zheng04}, who report a range of at least 2.5 dex in $\eta$
at $z\lesssim 3$. This may indicate the presence of local \HI-ionising
sources.

\begin{figure*}
\includegraphics{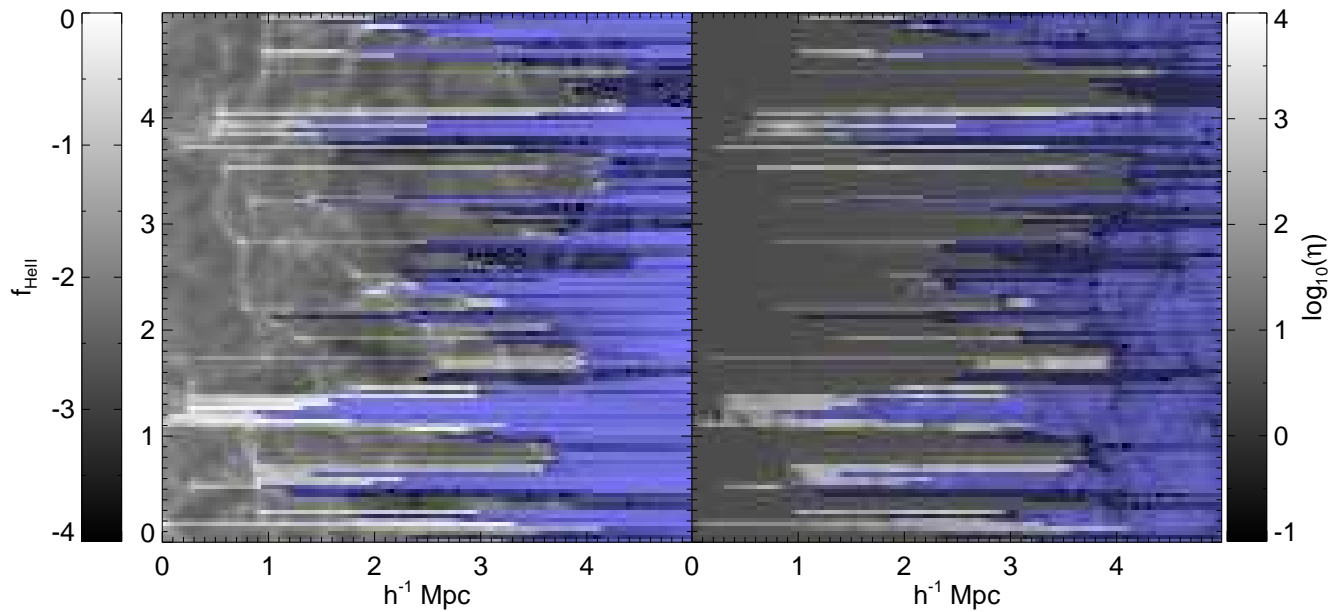}
\caption{Comparison of the \HeII\ ionisation fraction (left panel)
and the ratio of \HeII\ to \HI\ number densities
$\eta=n_\HeIIs/n_\HIs$ (right panel), for the hybrid model
HY20 at $z=4$. (Regions for which hydrogen is not ionised by $z=6$
are masked as blue.) A wide range in $\eta$ is found. The largest values
($\eta>1000$) correspond to regions where \HeII\ is still incompletely
reionised, with $f_\HeIIs \lsim 1$. The \HI\ fractions have been corrected
to a uniform photoionisation rate (see text).}
\label{fig.f_HeII-eta}
\end{figure*}

\citet{Zheng04} and \citet{Reimers05} report some regions with $\eta>1000$, suggesting
these may be regions for which \HeII\ is still not ionised to \HeIII.
\Fig{fig.f_HeII-eta} is a map of $\eta$ for the hybrid
model HY20 at $z=4$ compared with a map of the $f_\HeIIs$ fraction. Large
$\eta$ values are found to correspond to regions of high \HeII\
fractions. The evolution of the
$\eta$ distribution for the hybrid model HY20 is illustrated in \fig{fig.eta_dist}. At $z\gsim4$, when
the power-law source is just taking over from the starburst in
magnitude, high excursions are found for $\eta$, with values reaching
up to 3000. These high $\eta$ regions are associated with incomplete
\HeII\ reionisation, where $f_\HeIIs \lesssim 1$. Lowering the
redshift at which the power-law source overtakes the starburst would
lower the redshift at which these high excursions occur to
$z\lesssim3$. These results suggest that \citet{Reimers05} may have
detected the \HeII\ reionisation epoch.

In their simulations, \citet{MF05} find a higher value of $\eta\sim 250$, but a smaller spread of only a factor of two to three, not including the low values of $\eta$ associated with self-shielding.  As discussed earlier, our single I-front simulations produce an over-abundance of shielded \HI\ gas at $z<6$, which we have eliminated by correcting the $n_\HIs$ values to a uniform \HI\ photoionisation rate. Their higher value of $\eta$ results partially from the soft spectrum adopted from \citet{HM96} for QSOs with a $\alpha=1.5$ source spectrum.

\begin{table*}
\caption{Fits to the relation $f_\HIs = \alpha (\rho / \bar{\rho})^\beta$
for the gas for which $f_\HIs < 0.1$ by $z=6$. The errors are to the fits
and not indicative of the distribution in the $\rho-f_\HIs$ plane.}
\begin{tabular}{|c|c|c|c|c|c|c|}
 & \multicolumn{2}{c|}{PL20} & \multicolumn{2}{c|}{MQ20} & \multicolumn{2}{c|}{HY20}  \\
\cline{2-7}
$z$ & $\alpha$ $[\expd{-4}]$ & $\beta$ & $\alpha$ $[\expd{-4}]$ & $\beta$ & $\alpha$ $[\expd{-4}]$ & $\beta$ \\
\hline
6 & $5.093\pm 0.003$ & $0.847\pm 0.001$ & $3.674\pm 0.001$ & $0.8044\pm 0.006$ & $2.014\pm 0.002$ & $0.803\pm 0.002$ \\
\hline
5 & $3.770\pm 0.002$ & $0.756\pm 0.001$ & $2.784\pm 0.001$ & $0.721\pm 0.001$ & $1.680\pm 0.001$ & $0.726\pm 0.001$ \\
\hline
4 & $2.800\pm 0.002$ & $0.661\pm 0.001$ & $2.083\pm 0.001$ & $0.643\pm 0.001$ & $1.8587\pm 0.001$ & $1.066\pm 0.001$ \\
\hline
3 & $2.283\pm 0.001$ & $0.6226\pm 0.0005$ & $1.7239\pm 0.0004$ & $0.6096\pm 0.0003$ & $2.022\pm 0.002$ & $0.904\pm 0.001$ \\
\hline
\label{table.FractionFits_fH1}
\end{tabular}
\end{table*}

\begin{table*}
\caption{Fits to the relation $f_\HeIIs = \alpha (\rho / \bar{\rho})^\beta$
for the gas for which $f_\HIs < 0.1$ by $z=6$. The errors are to the fits
and not indicative of the distribution in the $\rho-f_\HIs$ plane.
Note there is negligible \HeII\ in the HY20 run prior to $z=5$.}
\begin{tabular}{|c|c|c|c|c|c|c|}
 & \multicolumn{2}{c|}{PL20} & \multicolumn{2}{c|}{MQ20} & \multicolumn{2}{c|}{HY20}  \\
\cline{2-7}
$z$ & $\alpha$ $[\expd{-2}]$ & $\beta$ & $\alpha$ $[\expd{-2}]$ & $\beta$ & $\alpha$ $[\expd{-2}]$ & $\beta$ \\
\hline
6 & $1.959\pm 0.002$ & $0.775\pm 0.001$ & $1.717\pm 0.002$ & $0.7508\pm 0.0009$ & & \\
\hline
5 & $1.531\pm 0.001$ & $0.695\pm 0.001$ & $1.3621\pm 0.0009$ & $0.6910\pm 0.0009$ & & \\
\hline
4 & $1.1989\pm 0.0006$ & $0.6464\pm 0.0007$ & $1.0514\pm 0.0004$ & $0.6380\pm 0.0005$ & $0.8657\pm 0.0008$ & $0.978\pm 0.001$ \\
\hline
3 & $1.0107\pm 0.0005$ & $0.6175\pm 0.0006$ & $0.8897\pm 0.0003$ & $0.6171\pm 0.0005$ & $0.9295\pm 0.0008$ & $0.847\pm 0.001$ \\
\hline
\label{table.FractionFits_fHe2}
\end{tabular}
\end{table*}

We parametrize the $f_{\rm HI}$ by the gas density,
$f_\HIs = \alpha(\rho/\bar{\rho})^\beta$, and similarly for $f_{\rm
HeII}$. The best fit values are provided in
Tables~\ref{table.FractionFits_fH1} and \ref{table.FractionFits_fHe2}.
Cosmic variance leads to errors in $\alpha$ of about 10 per cent and
$\beta$ of about 2 per cent. Although the values of $\alpha$ may be
readjusted by changing the magnitude of the ionisation radiation
background, the values of $\beta$ are invariant for a given source spectrum. The close agreement
between the values of $\beta$ for $f_\HIs$ and $f_\HeIIs$ for the
respective models reflects the near linear dependence between the \HI\ 
and \HeII\ ionisation fractions, with a weak residual dependence on
density (cf \fig{fig.f_h1_vs_f_he2}).

The power-law dependence of ionisation fraction on density is expected
if the shape of the local ionising spectrum is constant and the
density and temperature are related through a power law. The equilibrium
rate equation for hydrogen is \hbox{$n_\HIs \Gamma_\HIs = n_e n_\HIIs
\alpha_\HIIs$}. Assuming almost complete ionisation, the
ionisation fraction is,
\begin{equation}
f_\HIs \simeq \frac{n_e \alpha_\HIIs}{\Gamma_\HIs}.
\end{equation}
We recall that the photoionisation rate, $\Gamma_\HIs$, is a function
of the shape of the local ionising spectrum. Assuming it is constant
and that ionisation is almost complete, the neutral fraction follows
$f_\HIs \propto \rho \alpha_\HIIs$. Over the temperature range of
interest, the recombination rate responds to the temperature as
$\alpha_\HIIs \propto T^{-0.69}$. If the density and temperature are
related polytropically, $T \propto \rho^{\gamma - 1}$, then $f_\HIs
\propto \rho^{1 - 0.69(\gamma - 1)}$. From our power-law fits in \tab{table.FractionFits_fH1},
the inferred values for $\gamma$ at $z=3$ are $1.547\pm 0.001$,
$1.566\pm 0.001$, and $1.139\pm 0.001$ for PL20, MQ20, and HY20
respectively. For \HeII\, the inferred values at $z=3$ are $\gamma =
1.554\pm 0.001$, $1.555\pm 0.001$, and $1.222\pm 0.001$ PL20, MQ20,
and HY20 respectively. The values of
$\gamma$ are nearly identical to those found by directly fitting the
$\rho-T$ relation in \sect{sec.TemperatureResults}.

\section{Discussion and Conclusions}
\label{sec.Discussion}
We have coupled a radiative transfer code based on a probabilistic
photon transmission algorithm to a Particle-Mesh $N$-body code in
order to study the sensitivity of the post-ionisation temperature of
the Intergalactic Medium on source spectrum.  We performed multiple
simulations with different spectra of ionising radiation: a power-law
($\propto \nu^{-0.5}$), miniquasar, starburst, and a time-varying
hybrid spectrum that evolves from a starburst spectrum to a power
law.

The power-law and miniquasar spectra produce almost identical
temperature distributions, owing to their similar shapes. A greater
difference of temperatures is found between the remaining models. The
mass-weighted mean gas temperatures at $z=3$ are 9000~K for the
starburst source, 15000~K for the power-law and miniquasar source, and 17000~K for the hybrid models. A larger difference is found between the power-law/miniquasar
and hybrid model temperatures from the polytropic fits to temperature and density. A fit to the
polytropic relation, $T = T_0 (\rho/\bar{\rho})^{\gamma-1}$, gives
$T_0 = 14600\pm 200$, $14500\pm 160$, $3515\pm 11$, $18600\pm 100$ and
$\gamma = 1.58\pm 0.01$, $1.57\pm 0.01$, $1.553\pm 0.002$, $1.134\pm
0.005$ for the power-law, miniquasar, starburst and hybrid models,
respectively, at $z=3$. The errors are formal fit errors. The highest temperatures are found in the hybrid model at the end of the transition from starburst to power-law spectra ($z\approx4$), at which time the temperatures span values up to $40,000\K$, as required by measurements of \CIV\ and \SiIV\ 
absorption systems in the \Lya\ forest \citep{RSWB96}.

The \HeIII\ I-front passing through \HII-dominated gas leads to a
larger jump in temperature than the \HII\ I-front passing through
\HeIII-dominated gas. Indeed, in the simulations with the
power-law spectrum the temperature increase is only about 10 per
cent when a trailing \HII\ I-front passes through gas in which helium
is fully ionised. The difference is explained by the decrease in the mean molecular weight when \HI\ is ionised. Hence, a significant temperature change
will not be produced around a hard source by the passage of a \HII\
I-front, in contrast to a soft source or a region in which hydrogen
was fully ionised prior to helium.

The post-ionisation temperature of the IGM may be used as a key
observable for identifying the nature of the sources of reionisation. While moderate
to high overdensity gas establishes an equilibrium temperature in
which photoionisation heating balances atomic radiative cooling
processes, the equilibrium time scale exceeds a Hubble time in lower
density regions, those that give rise to optically thin \Lya\
absorption systems. In contrast to optically thin reionisation models, we find a
broad fanning out of the temperature-density relation for underdense
regions, with temperatures exceeding $3\times10^4$~K for the hybrid model with late \HeII\ reionisation. This may
help substantially in reconciling the much larger Doppler parameters
measured, which have a median value of about $30\,{\rm km\,s^{-1}}$ for the
optically thin \Lya\ systems at $z\approx3$, with those predicted by
simulations without radiative transfer \citep{MBM01}. Even allowing
for bulk motions to contribute as much as half to the line widths (in
quadrature), the large measured median Doppler parameters require a gas temperature of $2.8\times10^4$~K. The
hybrid model comes closest to meeting
this requirement. Ultimately, incorporating hydrodynamics is necessary
for definitive predictions of the detailed post-ionisation state of
the IGM and a precise determination of the resulting statistical
properties of the \Lya\ forest.

We do not address the effect of the epoch of reionisation on the
evolution of the gas temperature. Much earlier reionisation redshifts
for the entire IGM will result in much cooler temperatures, primarily
due to intense Compton cooling at high redshift. The effect of the
reionisation epoch on the subsequent IGM temperature has been explored
by \citet{IO86}, \citet{GS96}, and \citet{Theuns02} without radiative
transfer. A degeneracy exists between the epoch of reionisation and
the nature of the sources on the temperature of the IGM. If the epoch
of reionisation were determined through some other means, as for
instance its detection in the radio by {\it LOFAR} or the {\it MWA}, the
subsequent temperature of the IGM, particularly as revealed by the
widths of optically thin \Lya\ forest absorbers, could then be used to
determine the nature of the sources.

Hardening of the spectrum due to passage through structures with high
\HI\ column densities produces fluctuations in the $f_\HeIIs / f_\HIs$
ratio in the shadowed regions behind \HeIII\ I-fronts prior
to complete \HeII\ reionisation.
A spread is indeed found in the data at $z\lsim3$
\citep{Zheng04, Reimers05}. The observed spread of about 2.5 dex,
however, exceeds that found in our simulations at $z=3$ by about a factor of
three. In particular, values of $\eta>1000$ are reported by
\citet{Zheng04} and \citet{Reimers05}. It may be that local sources are required to
reproduce the wider range of observed values. Alternatively, at
$z\gsim4$ values up to $\eta\approx3000$ are found for the hybrid
model, when the power-law source begins to dominate over the starburst
producing partially ionised \HeII.\footnote{A similar effect is likely
to occur for a power-law spectrum with $\alpha>1.8$, as in this case
the helium I-front lags behind the hydrogen I-front.} Lowering the redshift
at which the power-law spectrum dominates over the starburst, thus delaying the
\HeII\ reionisation epoch, would lower the redshift at which large fluctuations
in $\eta$ are produced. It may be that
\citet{Reimers05} have detected the epoch of \HeII\ reionisation.

Not explored in detail by this paper is the pre-ionisation state of the gas which is relevant to future $21\cm$ detections.  To correctly model the gas will require the incorporation of the production of secondary electrons and the diffuse radiation field. The production of the secondary electrons further cools the gas due both to overcoming the ionisation potential and to radiative losses from the consequent enhancement of \Lya\ collisional excitation \citep{Shull79}. A copious number of \Lya\ photons could be produced in any region where helium is ionised prior to hydrogen. The photons may have important implications for the detection of the 21cm signature of the IGM, as they can decouple the spin temperature from the CMB \citep{MMR97}. The \Lya\ photons will also be an additional source of pre-heating through recoils, although the amount is likely to be small before the scattering reaches equilibrium \citep{Meiksin06}. We are currently including both hydrodynamics and the extra radiative physics in more refined models. The addition of hydrodynamics will also eliminate any limitations owing to the absence of advection.

\section*{Acknowledgements}
Some of the computations reported here were performed using the UK
Astrophysical Fluids Facility (UKAFF). Tittley is supported by a
PPARC Rolling-Grant.

\bibliographystyle{mn2e-eprint}
\bibliography{apj-jour,biblio}

\label{lastpage}

\end{document}